\documentclass{aastex63}

\usepackage{amsmath}
\usepackage{CJK}
\usepackage{float}
\usepackage[normalem]{ulem}
\usepackage[shortlabels]{enumitem}
\usepackage{multirow}
\definecolor{myred}{RGB}{255,50,50} 
\definecolor{myblue}{RGB}{0,0,180}    
\definecolor{mygrey}{RGB}{184,134,11}  
\definecolor{green}{RGB}{0,180,0}


\def\Gaia  {$Gaia$}
\def\hho     {H$_2$O}
\def\VLSR    {$V_{\rm LSR}$}

\def\kms     {km~s$^{-1}$}
\def\masy    {mas~yr$^{-1}$}

\def\mjybeam {mJy~beam$^{-1}$}
\def\jybeam  {Jy~beam$^{-1}$}

\def\uas     {$\mu$as}

\def\ra      {$\alpha_{\rm J2000}$}
\def\dec     {$\delta_{\rm J2000}$}
\def\h       {\ifmmode{^{\rm h}}\else{$^{\rm h}$}\fi}
\def\m       {\ifmmode{^{\rm m}}\else{$^{\rm m}$}\fi}
\def\s       {\ifmmode{^{\rm s}}\else{$^{\rm s}$}\fi}
\def\deg     {\ifmmode{^{\circ}}\else{$^{\circ}$}\fi}
\def\decdeg  {\ifmmode{{\rlap.}^{\circ}} \else ${\rlap.}^{\circ}$\fi}
\def\decs    {\ifmmode{{\rlap.}^{\rm s}} \else ${\rlap.}^{\rm s}$\fi}
\def\decas   {\ifmmode{{\rlap.}{''}}\else{${\rlap.}{''}$}\fi}
\def\Ho  {\ifmmode{{\rm H}_0}\else{H$_0$}\fi}
\def\Ro  {\ifmmode{{\rm R}_0}\else{R$_0$}\fi}
\def\To  {\ifmmode{\Theta_0}\else{$\Theta_0$}\fi}

\def\Vsbar {\ifmmode {\overline{V_s}}\else {$\overline{V_s}$}\fi}
\def\Usbar {\ifmmode {\overline{U_s}}\else {$\overline{U_s}$}\fi}
\def\Wsbar {\ifmmode {\overline{W_s}}\else {$\overline{W_s}$}\fi}

\def\mux    {\ifmmode {\mu_x}\else {$\mu_x$}\fi}
\def\muy    {\ifmmode {\mu_y}\else {$\mu_y$}\fi}
\def\mura   {\ifmmode {\mu_{\alpha}}\else {$\mu_{\alpha}$}\fi}
\def\mude   {\ifmmode {\mu_{\delta}}\else {$\mu_{\delta}$}\fi}

\def\gax{\mathrel{\rlap{\lower4pt\hbox{\hskip1pt$\sim$}}
    \raise1pt\hbox{$>$}}}

\def\d    {\ifmmode {{\rlap{.}}^\circ}\else {${\rlap{.}}^\circ$}\fi}
\def\s    {\ifmmode {{\rlap{.}}^s}\else {${\rlap{.}}^s$}\fi}
\def\as   {\ifmmode {{\rlap{.}}^{''}}\else {${\rlap{.}}^{''}$}\fi}

\def\uas {$\mu$as}
\def\masy{mas~yr$^{-1}$}

\def\kms {km~s$^{-1}$}

\def\vlsr{$V_{\mbox{\scriptsize LSR}}$}
\def\kms{~km~s$^{-1}$}
\def\kmsyr{~km~s$^{-1}$yr$^{-1}$}

\def\hho{H$_{2}$O}

\def\lax{\mathrel{\rlap{\lower4pt\hbox{\hskip1pt$\sim$}}
    \raise1pt\hbox{$<$}}}                % less than or approx. symbol
\def\gax{\mathrel{\rlap{\lower4pt\hbox{\hskip1pt$\sim$}}
    \raise1pt\hbox{$>$}}}                % greater than or approx. symbol
    
\def\aap{A\&A}

\def\apjs{ApJS}
\def\apjl{ApJ}

\shorttitle{Animation of \hho\ masers around BX Cam}
\shortauthors{Xu et al.}
\graphicspath{{./}{figures/}}

\begin{document}
\begin{CJK*}{UTF8}{gbsn}
\title{The Astrometric Animation of Water Masers towards the Mira Variable BX Cam}

\correspondingauthor{Shuangjing Xu}
\email{sjxuvlbi@gmail.com}

\author[0000-0003-2953-6442]{Shuangjing Xu %(徐双敬)
} 
\affiliation{Korea Astronomy and Space Science Institute, 776 Daedeok-daero, Yuseong-gu, Daejeon 34055, Republic of Korea}
\affiliation{Shanghai Astronomical Observatory, Chinese Academy of Sciences, 80 Nandan Road, Shanghai 200030, China}
\author[0000-0002-0880-0091]{Hiroshi Imai %(今井裕)
}
\affiliation{Graduate School of Science and Engineering, 
Kagoshima University, \\
1-21-35 Korimoto, Kagoshima 890-0065, Japan}
\affiliation{Amanogawa Galaxy Astronomy Research Center, Graduate School of Science and Engineering, Kagoshima University,  \\
1-21-35 Korimoto, Kagoshima 890-0065, Japan}
\affiliation{Center for General Education, Institute for Comprehensive Education, Kagoshima University,  \\
1-21-30 Korimoto, Kagoshima 890-0065, Japan}

\author{Youngjoo Yun} 
\affiliation{Korea Astronomy and Space Science Institute, 776 Daedeok-daero, Yuseong-gu, Daejeon 34055, Republic of Korea}

\author{Bo Zhang %(张波)
}
\affiliation{Shanghai Astronomical Observatory, Chinese Academy of Sciences, 80 Nandan Road, Shanghai 200030, China}

\author[0000-0003-4871-9535]{Mar\'{\i}a J. Rioja}
\affiliation{ICRAR, M468, The University of Western Australia, 35 Stirling Hwy, Crawley, Western Australia, 6009, Australia}
\affiliation{CSIRO Astronomy and Space Science, PO Box 1130, Bentley WA 6102, Australia}
\affiliation{Observatorio Astron\'omico Nacional (IGN), Alfonso XII, 3 y 5, 28014 Madrid, Spain}

\author[0000-0003-0392-3604]{Richard Dodson}
\affiliation{ICRAR, M468, The University of Western Australia, 35 Stirling Hwy, Crawley, Western Australia, 6009, Australia}

\author{Se-Hyung Cho} 
\affiliation{Astronomy program, Department of Physics and Astronomy, Seoul National University, Seoul 08826, Republic of Korea/Korea Astronomy and Space Science Institute, Yuseong–gu, Daejeon 34055, Republic of Korea}

\author[0000-0001-9825-7864]{Jaeheon Kim} 
\affiliation{Korea Astronomy and Space Science Institute, 776 Daedeok-daero, Yuseong-gu, Daejeon 34055, Republic of Korea}

\author[0000-0003-0721-5509]{Lang Cui %(崔朗) 
} 
\affiliation{Xinjiang Astronomical Observatory, Chinese Academy of Sciences, 150 Science 1-Street, Urumqi 830011, China}

\author[0000-0001-7575-5254]{Andrey M. Sobolev}
\affiliation{Ural Federal University, 19 Mira Street, 620002 Ekaterinburg, Russia}

\author[0000-0002-9875-7436]{James O. Chibueze}
\affiliation{Centre for Space Research, North-West University, Potchefstroom 2520, South Africa}
\affiliation{Department of Physics and Astronomy, Faculty of Physical Sciences,\\
University of Nigeria, Carver Building, 1 University Road, Nsukka, Nigeria}

\author{Dong-Jin Kim}
\affiliation{Massachusetts Institute of Technology Haystack Observatory, 99 Millstone Road, Westford, MA 01886, USA}
\affiliation{Max-Planck-Institut für Radioastronomie, Auf dem Hügel 69, D-53121 Bonn, Germany}

\author{Kei Amada}
\affiliation{Graduate School of Science and Engineering, 
Kagoshima University, \\
1-21-35 Korimoto, Kagoshima 890-0065, Japan}

\author{Jun-ichi Nakashima}
\affiliation{School of Physics and Astronomy, Sun Yat-sen University, 2 Daxue Road, Tangjia, Zhuhai, Guangdong Province,  China}

\author{Gabor Orosz} 
\affiliation{Joint Institute for VLBI ERIC, Oude Hoogeveensedijk 4, 7991PD Dwingeloo, Netherlands}

\author{Miyako Oyadomari}
\affiliation{Graduate School of Science and Engineering, 
Kagoshima University, \\
1-21-35 Korimoto, Kagoshima 890-0065, Japan}

\author{Sejin Oh} 
\affiliation{Korea Astronomy and Space Science Institute, 776 Daedeok-daero, Yuseong-gu, Daejeon 34055, Republic of Korea}

\author[0000-0001-5615-5464]{Yoshinori Yonekura} 
\affiliation{Center for Astronomy, Ibaraki University, 2-1-1 Bunkyo, Mito, Ibaraki 310-8512, Japan}

\author{Yan Sun}
\affiliation{Shanghai Astronomical Observatory, Chinese Academy of Sciences, 80 Nandan Road, Shanghai 200030, China}
\affiliation{University of Chinese Academy of Sciences, 
No.19 (A) Yuquan Rd. Shijingshan, Beijing, 100049, China}

\author{Xiaofeng Mai}
\affiliation{Shanghai Astronomical Observatory, Chinese Academy of Sciences, 80 Nandan Road, Shanghai 200030, China}
\affiliation{University of Chinese Academy of Sciences, 
No.19 (A) Yuquan Rd. Shijingshan, Beijing, 100049, China}

\author{Jingdong Zhang}
\affiliation{Shanghai Astronomical Observatory, Chinese Academy of Sciences, 80 Nandan Road, Shanghai 200030, China}
\affiliation{University of Chinese Academy of Sciences, 
No.19 (A) Yuquan Rd. Shijingshan, Beijing, 100049, China}

\author{Shiming Wen}
\affiliation{Shanghai Astronomical Observatory, Chinese Academy of Sciences, 80 Nandan Road, Shanghai 200030, China}

\author{Taehyun Jung} 
\affiliation{Korea Astronomy and Space Science Institute, 776 Daedeok-daero, Yuseong-gu, Daejeon 34055, Republic of Korea}

\begin{abstract} 

We report VLBI monitoring observations of the 22\,GHz water (\hho) masers around the Mira variable BX Cam, which were carried out as a part of the EAVN Synthesis of Stellar Maser Animations (ESTEMA) project. Data of 37 epochs in total were obtained from 2018 May to 2021 June with a time interval of 3--4 weeks, spanning approximately three stellar pulsation periods ($P= \sim$440\,d). In particular, the dual-beam system equipped on the VERA stations was used to measure the kinematics and parallaxes of the \hho\ maser features. The measured parallax, $\pi=1.79\pm 0.08$~mas, is consistent with \Gaia\ EDR3 and previously measured VLBI parallaxes within a 1-$\sigma$ error level. The position of the central star was estimated, based on both the \Gaia\ EDR3 data and the center position of the ring-like 43\,GHz silicon-monoxide (SiO) maser distribution imaged with the KVN. The three-dimensional \hho\ maser kinematics indicates that the circumstellar envelope is expanding at a velocity of $13\pm4$ \kms, while there are asymmetries in both the spatial and velocity distributions of the maser features. Furthermore, the \hho\ maser animation achieved by our dense monitoring program manifests the propagation of shock waves in the circumstellar envelope of BX Cam.

\end{abstract}


\section{Introduction} \label{sec:intro}

The long-period variables (LPVs), such as Mira variables, are stars of low to intermediate masses which have reached the late evolutionary stage of Asymptotic Giant Branch (AGB) phase. They are characterised by 
long-period ($>$100 d) variations in radius, brightness, and temperature, which are caused by stellar surface instability and radial pulsation  (see the latest catalogue of  
\citealt{2022arXiv220605745L}). These stars also have an  intense mass-loss phenomenon or a superwind, which leads to  formation of a circumstellar envelope (CSE) made of gas and dust (see the review of \citealt{2018A&ARv..26....1H}). 

The stellar mass loss may be driven, primarily, by the pressure from the stellar radiation on dust grains, while pulsation-induced shocks are expected to enhance the mass loss (e.g., \citealt{2004agbs.book.....H,2011ASPC..445..193H,2016A&A...594A.108H}).
In addition, 
the stellar structure is basically spherically symmetric, 
while the mass loss often exhibits significant asymmetry 
on very different spatial (1--10$^4$ AU) and temporal (a few months to a few 10$^4$ years) scales  \citep{2018A&ARv..26....1H}. The mass loss asymmetry observed in the vicinity of the stellar surface (e.g., \citealt{2020A&A...635A.200K}) might be linked to the existence of giant gas segments created from convection in the stellar interior. It is interesting to trace how such asymmetry will grow while the gas segments are being accelerated outward. It is likely that the mechanism of dust acceleration in the CSE, the enhancement of the outward flow, and the growth of its asymmetry may be dependent on the property or abundance of the ejected materials (i.e., carbon-rich or oxygen-rich composition) \citep{2018A&ARv..26....1H}.

Circumstellar SiO and \hho\ masers are located at, respectively, $\sim$2-4 stellar radii (e.g., \citealt{1994ApJ...430L..61D}) and $\sim$5-50 stellar radii (e.g., \citealt{2012A&A...546A..16R}) of a typical oxygen-rich AGB star, while the dust formation layer is co-located between the SiO masers and  \hho\ masers \citep{1996A&ARv...7...97H,2007A&A...470..191W}.
The location of SiO masers corresponds to the innermost part of the CSE, where molecules are still rich and outward and inward motions are often observed  (e.g. \citealt{2013MNRAS.433.3133G}). 
The location of \hho\ masers corresponds to the zone of the outward acceleration of dust and gas, where molecular gas is cooled and condensed to dust while \hho\ molecules are still enriched. The maser structures are composed of clusters of compact maser features (isolated gas clumps with sizes of $\sim$1~AU). This enables us to measure the three-dimensional velocity field of the CSE through monitoring observations by Very Long Baseline Interferometry (VLBI) with excellent angular resolution (e.g., \citealt{2003ApJ...590..460I}). Groups of maser features may trace giant gas/dust clumps, whose origins have been debated in terms of the convection of the stellar interior or just instability on the stellar surface \citep{2018A&ARv..26....1H}. Thus, the maser spatio-kinematics have direct relationship to such gas/dust dynamics and can illuminate important issues on the mechanisms of maser excitation. 

The EAVN Synthesis of Stellar Maser Animations (ESTEMA) project has been conducted as one of the KaVA (a combined array of the Korean VLBI Network, KVN, and Japanese VLBI Exploration of Radio Astrometry, VERA) Large Program projects followed by a series of General Proposal projects of the East Asian VLBI Network (EAVN), aiming at intensive VLBI monitoring observations of CSE masers associated with LPVs of different pulsation periods (300---1000~d) over a few stellar pulsation cycles. EAVN is suitable for high cadence VLBI monitoring because more than six telescopes are always available for 
imaging of CSE masers. 
The typical angular resolutions are 1.2(K)/0.6(Q) mas  with the core array KaVA, and the largest detectable angular scales are 9(K)/5(Q) mas with the short baselines of $\sim$300 km provided by KVN. The expected image sensitivities are $\sim$30(K)/50(Q)  \mjybeam\ for KaVA  and  $\sim$40(K)/60(Q)  \mjybeam\ for KVN, with an integration time of 4 hours and narrower bandwidth of 15.625 kHz for maser emission.
In particular, it enables us to simultaneously monitor four to eight lines of \hho\ and SiO masers at four frequency (22/43/86/129~GHz or K/Q/W/D) bands with the KVN. Absolute positions, proper motions of maser gas clumps, and the annual parallax of the maser source system can be determined by precise astrometry with the dual-beam system of VERA. The animation of circumstellar masers  would demonstrate in detail the behaviors of maser emission,
reflecting the dynamics of the maser clumps and variation in regions of maser excitation around the LPVs. 

Here we present the first result of ESTEMA, which has been derived from \hho\ masers associated with the Mira variable star BX Camelopardalis (BX Cam). This star is an oxygen-rich Mira variable with 
similar, but slightly inconsistent, reported
pulsating periods of 486~d (AAVSO)\footnote{\url{https://www.aavso.org}} or 440~d \citep{2020PASJ...72...56M}. The spectral type is M9.5 \citep{1978A&AS...34..409S}, and the presence of bright SiO, \hho, and OH masers have been confirmed \citep{1984A&A...138..343U,2020PASJ...72...56M}.
The trigonometric parallax and the spatial distribution of \hho\ masers with three collimated flows have been reported by \cite{2020PASJ...72...56M}.
This star was selected as one of the ESTEMA targets because of the more flexible allocation of VLBI observation sessions for this Northern-sky star.
Similarly to the intensive monitoring observations of SiO ({\it $v$=1, J=1$\rightarrow$0}) masers around TX Cam with the Very Long Baseline
Array (VLBA) \citep{2003ApJ...599.1372D,2010MNRAS.406..395G,2013MNRAS.433.3133G}, the present monitoring project enables us to elucidate the global behaviour of an example of CSE \hho\  masers for the first time. CSE \hho\ masers also should be physically related to the stellar pulsation because of the periodic variation in the maser spectra  \citep{2008PASJ...60.1077S,2014AJ....147...22K}. However, the true origin of such a maser behavior can be visualized by an animation of the VLBI images.  The astrometry on the maser animation is crucial to exactly trace the motions of the individual maser features. After subtracting maser position modulations due to the annual parallax and the secular motion of the maser source system, the maser maps can be registered on the reference frame, which then provides a fixed frame that can be compared with the location of the pulsating star whose astrometric data is now available from the \Gaia\ Early Data Release 3 (\Gaia\ EDR3) \citep{2021A&A...649A...1G}. 
A maser animation derived based on above astrometric steps
enable us to visualize a radial expansion of the CSE masers and some deviations of the maser motions from constant-velocity motions, such as radial accelerations, rotations, etc.

\section{Observations and data Reduction}
The monitoring observations of \hho\ and SiO masers around BX Cam in ESTEMA have been ongoing since 2018 May, except in the EAVN maintenance season (June--August). The EAVN telescopes that have participated in ESTEMA observations include four telescopes of VERA (Mizusawa, Iriki, Ogasawara, and Ishigakijima), three telescopes of the KVN (Yonsei, Ulsan, Tamna), two telescopes of the Chinese VLBI Network (CVN: Tianma, Urumqi in K-band only), and one Japanese telescope (Takahagi in K-band only). The time cadence for a single observational epoch is 3--4 weeks, which is approximately equal to a time resolution of 1/20 stellar pulsation cycles of BX~Cam ($\sim$22~d). Each observation epoch corresponds to a pair of K and Q-band experiments over two consecutive days and lasts for 8~hr (2~d$\times$4~hr). We have designed the observations to use all of the unique capabilities of the KVN and VERA in EAVN, using a hybrid setup of simultaneously monitoring  \hho\ (22.235080 GHz) and SiO (42.820539, 43.122027, 86.243442, and 129.363358 GHz) maser lines and conducting high-accuracy astrometry by observing the maser and the reference continuum sources simultaneously.

In this paper, we mainly focus on the K-band data as listed in Table \ref{tab:obs}, spanning observations from 2018 to 2021. The data sample has crossed three light curve maxima of the stellar pulsation from phase $\phi$ 0.65 to $\phi$  3.17.
Each of the observation sessions had a duration of 4~hr including scans on the target, BX Cam, the fringe finder, NRAO150, and the delay calibrator, J0721+7120 (\ra\ = 07\h21\m53\decs4485$\pm$0.31 mas, \dec\ = 71\degr20\arcmin 36\decas363$\pm$0.10 mas).
We alternated scans between BX Cam for 7~min and the calibrator for 2~min. The angular separation between these two sources is 7.9\deg\ on the sky, which is suitable for source-frequency phase-referencing (SFPR) analysis with KVN data \citep{2014AJ....148...97D}. 
Meanwhile,  the phase reference source J0524$+$7034 (\ra\ = 05\h24\m13\decs4333$\pm$0.39 mas, \dec\ = 70\degr34\arcmin 52\decas906$\pm$0.13 mas \footnote{\label{note1}
The quasar positions are cited from \url{http://astrogeo.org/sol/rfc/rfc_2021d/}.}), 2.0\deg\ away from the target, was observed by the second beam of VERA to determine the absolute positions of the maser spots of BX Cam.  

The received signals in left circular polarization were recorded in all the EAVN stations with 16 base-band channels (BBCs) each with a bandwidth of 16 MHz, one of which includes the maser emission of BX Cam. In VERA, 15 BBCs were assigned to the signals received by the second beam so as to cover a total width of 496 MHz. The recorded signal data were correlated using three passes with the hardware correlator in Daejeon, Korea \citep{2015JKAS...48..125L}. 
The first pass was processed for the BBC including \hho\ masers with all EAVN antennas and yielded 2048 spectral channels, corresponding to a velocity spacing of 0.105\kms\ at 22.235 GHz. The second pass was processed for the phase reference source J0524$+$7034 observed by the second beam of VERA. 
The third pass was for simultaneously monitoring of \hho\ and the four SiO maser lines using the KVN. In this pass, all of the 16 BBCs were processed with 1024 spectral channels per BBC.

The data calibration was performed with the Astronomical Image Processing System (AIPS) \citep{2003ASSL..285..109G} developed by the National Radio Astronomy Observatory (NRAO). We have developed the pipeline scripts with ParselTongue \citep{2006ASPC..351..497K} for ESTEMA project. A priori amplitude calibration was done using the system noise temperatures and antenna gains logged during observations at KVN and VERA stations.  
These system noise temperatures were measured with the chopper-wheel method \citep{1976ApJS...30..247U} and were evaluated using the “R-Sky” method \citep{2004PASJ...56L..15H}, by observing a reference black body. 
Since the system temperature measurements were absent for Tianma, Urumqi, and Takahagi stations,  
we applied the template spectrum method  \citep{1999ASPC..180..481R}, in which a template spectrum of the \hho\ maser emission was obtained from the auto-correlation data. The estimated amplitude calibration errors of both methods were less than 10$\%$ for EAVN \citep{2017PASJ...69...87C}.
For imaging the \hho\ masers with all the EAVN antennas,
the velocity channel including a bright, single, and compact maser spot was selected as the reference for fringe fitting and self-calibration.
The detection limit was typically 0.4 Jy beam$^{-1}$ at a 5-$\sigma$ noise 
level, in maps without bright maser emission. 

For astrometry with VERA, 
the instrumental relative delays between the dual beams measured by the horn-on-dish method are applied \citep{2008PASJ...60..935H} 
and the delay re-calculation tables \citep{2020PASJ...72...52N} were used. In the delay re-calculation tables, the more accurate delay models were updated, with: the tropospheric zenith delay measured by the Global Positioning System (GPS), the ionospheric Total Electron Content (TEC) of the Global Ionosphere Map (GIM) produced by the Center for Orbit Determination in Europe (CODE), the station position measured in the monthly geodetic VLBI observations, and the updated maser positions  \citep[and the references therein]{2020PASJ...72...52N}. 
Then we used J0524$+$7034 as the phase reference source to determine the absolute position of masers. 
5 $\sigma$ noise levels of the phase referenced map of VERA data are typically 1.0 Jy beam$^{-1}$.
Therefore, we can measure the absolute positions of the bright masers of BX Cam with the VERA dual-beam bona-fide astrometry,
then propagate this to all of the maser spots in the EAVN image by registering the bright masers in the pair of images.
We used the SFPR technique described in \cite{2014AJ....148...97D} to obtain the astrometric registration of the \hho\ and SiO maser spots from the KVN data.

\begin{deluxetable*}{lrrllrrcrrrr}
\tablenum{1}
\tablecaption{EAVN observations of BX~Cam at K band  \label{tab:obs}}
\tablewidth{0pt}
\tablehead{
\colhead{Obs.}  & \colhead{Epoch}  & \colhead{MJD\tablenotemark{a}}  & Optical  & \colhead{Stations\tablenotemark{c}} & \colhead{Num.\tablenotemark{d}} & \colhead{RMS\tablenotemark{e}} &\colhead{High-Res.\tablenotemark{d}}  & \colhead{VERA\tablenotemark{d}}  & \colhead{Problem\tablenotemark{f}} & \nocolhead{Problem}\\
\colhead{Code} & \colhead{}  & \colhead{} &  \colhead{Phase($\phi$)\tablenotemark{b}} & \colhead{} & \colhead{} & \colhead{}  & \colhead{Imaging} & \colhead{Astrometry} & 
}
\startdata
 k18hi01a   & 2018-05-24 & 58262 & 0.65 & VmVrVoVsKyKuKt            & 1  & 283   &            &             &      & \\
 k18hi01c   & 2018-06-06 & 58275 & 0.68 & VmVrVoVsKyKuKt            & 1  & 320   &            &             &     & \\
 k18hi01e   & 2018-09-03 & 58364 & 0.88 & VmVrVoVs~~~~KuKt          & 2  & 278   &            &             &  & \\
 k18hi01g   & 2018-10-07 & 58398 & 0.96 & ~~~~~VrVoVsKyKuKt         & 0  & 1513  &            &             & Kt(N), KyKu(R)  &  \\
 k18hi01i   & 2018-10-24 & 58415 & 1.00 & VmVrVoVsKyKuKt~~~~Ur      & 13 & 210   & \checkmark &             &  KyKuKt(R) & \\
 k18hi01k   & 2018-11-11 & 58433 & 1.04 & VmVrVoVsKyKuKtT6Ur        & 16 & 122   & \checkmark & \checkmark  & T6Ur(N)  &\\
 k18hi01m   & 2018-12-16 & 58468 & 1.12 & VmVrVoVsKyKuKt            & 21 & 83    & \checkmark & \checkmark  &     & \\
 k18hi01o   & 2019-01-17 & 58500 & 1.19 & VmVrVoVsKyKuKtT6          & 15 & 62    & \checkmark & \checkmark  &    & \\
 k19hi01a   & 2019-02-11 & 58525 & 1.25 & VmVrVoVsKyKuKt            & 17 &  89   & \checkmark &             & Vm(N) & \\
 k19hi01c   & 2019-03-14 & 58556 & 1.32 & VmVrVoVsKyKuKt            & 12 &  96   &            &             &      & \\
 k19hi01e   & 2019-04-11 & 58584 & 1.38 & VmVrVoVsKyKuKt            & 5  &  123  &            &             &      & \\
 k19hi01g   & 2019-05-08 & 58611 & 1.44 & VmVrVo~~~~KyKuKt          & 7  & 98    &            &             &     & \\
 a19hi03a   & 2019-09-07 & 58733 & 1.72 & VmVr~~~~Vs~~~~KuKtT6      & 0  & 1461  &            &             & T6(N)  \\
 a19hi03c   & 2019-10-10 & 58766 & 1.79 & VmVr~~~~VsKy~~~~KtT6      & 6  & 154   & \checkmark &             & VmKy(N)  &\\
 a19hi03e   & 2019-10-24 & 58780 & 1.83 & VmVr~~~~VsKyKuKt          & 19 & 128   & \checkmark &             &    & \\
 a19hi03g   & 2019-11-18 & 58805 & 1.88 & VmVrVoVsKyKuKtT6          & 23 &  61   & \checkmark & \checkmark  &     &\\
 a19hi03i   & 2019-12-16 & 58833 & 1.95 & VmVrVoVsKyKuKtT6          & 32 &  65   & \checkmark & \checkmark  &    & \\
 a19hi03l   & 2020-01-10 & 58858 & 2.00 & VmVrVoVsKyKuKtT6          & 35 & 52    & \checkmark & \checkmark  &    & \\
 a2004a     & 2020-02-07 & 58886 & 2.07 & VmVrVoVsKyKuKt            & 38 &  57   & \checkmark & \checkmark  &     & \\
 a2004c     & 2020-03-11 & 58919 & 2.14 & VmVrVoVsKyKuKt            & 28 &   53  & \checkmark & \checkmark  &     & \\
 a2004e     & 2020-03-27 & 58935 & 2.18 & VmVrVoVs~~~~KuKtT6Ur      & 16 &  97   & \checkmark & \checkmark  &     & \\
 a2004g     & 2020-04-20 & 58959 & 2.23 & VmVrVoVsKyKuKt~~~~Ur      & 9  & 94    & \checkmark & \checkmark  & Ku(N)  &\\
 a2004i     & 2020-05-13 & 58982 & 2.29 & VmVrVoVsKy~~~~Kt~~~~Ur    & 12 &  88   & \checkmark & \checkmark  &     & \\
 a2004k     & 2020-06-09 & 59009 & 2.35 & VmVrVo~~~~KyKuKt~~~~Ur    & 11 &  136  &            &             &      \\
 a2019a     & 2020-09-15 & 59107 & 2.57 & Vm~~~~VoVsKyKu~~~~~~~Ur   & 0  & 765   &            &             & Vs(N) \\
 a2019c     & 2020-09-24 & 59116 & 2.59 & VmVrVoVsKyKuKt~~~~Ur      & 0  & 456   &            &             &       & \\
 a2019e     & 2020-10-18 & 59140 & 2.64 & ~~~~~VrVoVsKyKuKt~~~~Ur   & 4  & 167   &            &             &      & \\
 a2019g     & 2020-11-12 & 59165 & 2.70 & ~~~~~VrVoVsKyKuKt         & 6  & 97    &            &             &      & \\
 a2019i     & 2020-11-26 & 59179 & 2.73 & VmVrVoVsKyKuKt~~~~~~~Tk   & 6  & 128   &            &             &     & \\
 a2019m     & 2021-01-12 & 59226 & 2.84 & VmVrVoVs~~~~KuKt~~~~~~~Tk & 7  & 93    &            &             &   & \\
 a2102a     & 2021-02-01 & 59246 & 2.89 & VmVrVoVsKyKuKt~~~~~~~Tk   & 15 & 68    & \checkmark &             &     & \\
 a2102c     & 2021-02-19 & 59264 & 2.93 & VmVrVoVsKyKuKt~~~~~~~Tk   & 21 & 51    & \checkmark & \checkmark  &     & \\
 a2102e     & 2021-03-12 & 59285 & 2.97 & VmVrVoVsKyKu~~~~~~~~~~~Tk & 7  & 183   & \checkmark & \checkmark  & KyKu(R)   & \\
 a2102g     & 2021-04-05 & 59309 & 3.03 & VmVrVoVsKy~~~~KtT6UrTk    & 9  & 176   & \checkmark & \checkmark  &  KyKt(R)  & \\
 a2102i     & 2021-04-17 & 59321 & 3.06 & VmVrVoVsKyKuKt~~~~~~~Tk   & 6  & 598   & \checkmark &             &    Ky(N), KuKt(R) &\\
 a2102k     & 2021-05-12 & 59346 & 3.11 & VmVrVoVsKyKuKt~~~~~~~Tk   & 14 & 439   & \checkmark &             &    KyKuKt(R)  & \\
 a2102m     & 2021-06-07 & 59372 & 3.17 & VmVrVoVsKyKuKtT6Ur        & 16 & 111   & \checkmark &             &    Ur(N)  & \\
\enddata
\tablenotetext{a}{MJD: Modified Julian Day.}
\tablenotetext{b}{The optical phase is estimated in Figure \ref{fig:curve} using a mean period 440 days, optical maximum at 2020-01-08 and an uncertainty about 4 days.}
\tablenotetext{c}{Station codes represent VERA-Mizusawa(Vm), VERA-Iriki(Vr), VERA-Ogasawara(Vo), VERA-Ishigakijima(Vs), KVN-Yonsei(Ky), KVN-Ulsan(Ku), KVN-Tamna(Kt), Tianma(T6), Urumqi(Ur), and Takahagi(Tk), respectively.}
\tablenotetext{d}{The purposes of the data sets: (1) ``Num.'': Number of the maser features identified. (2)``High-Res. Imaging'':  The images made including  $>$2000 km baselines, which are used for analysing  the maser features and relative proper motions.
(3)``VERA Astrometry'': Phase-referencing astrometry with VERA dual-beam system, which are used for determining the absolute positions, absolute proper motions and annual parallaxes.}
\tablenotetext{e}{Noise level (1-$sigma$) of EAVN images in \mjybeam.}
\tablenotetext{f}{(N): No fringe detection. (R) RCP: polarization fault due to wrong observation in right-hand circular polarization.}
\end{deluxetable*}

\section{Results}

\subsection{Periodic variation of the \texorpdfstring{\hho}{H2O} maser spectra}
\label{sec:spec1}

\begin{figure*}[t]
\plotone{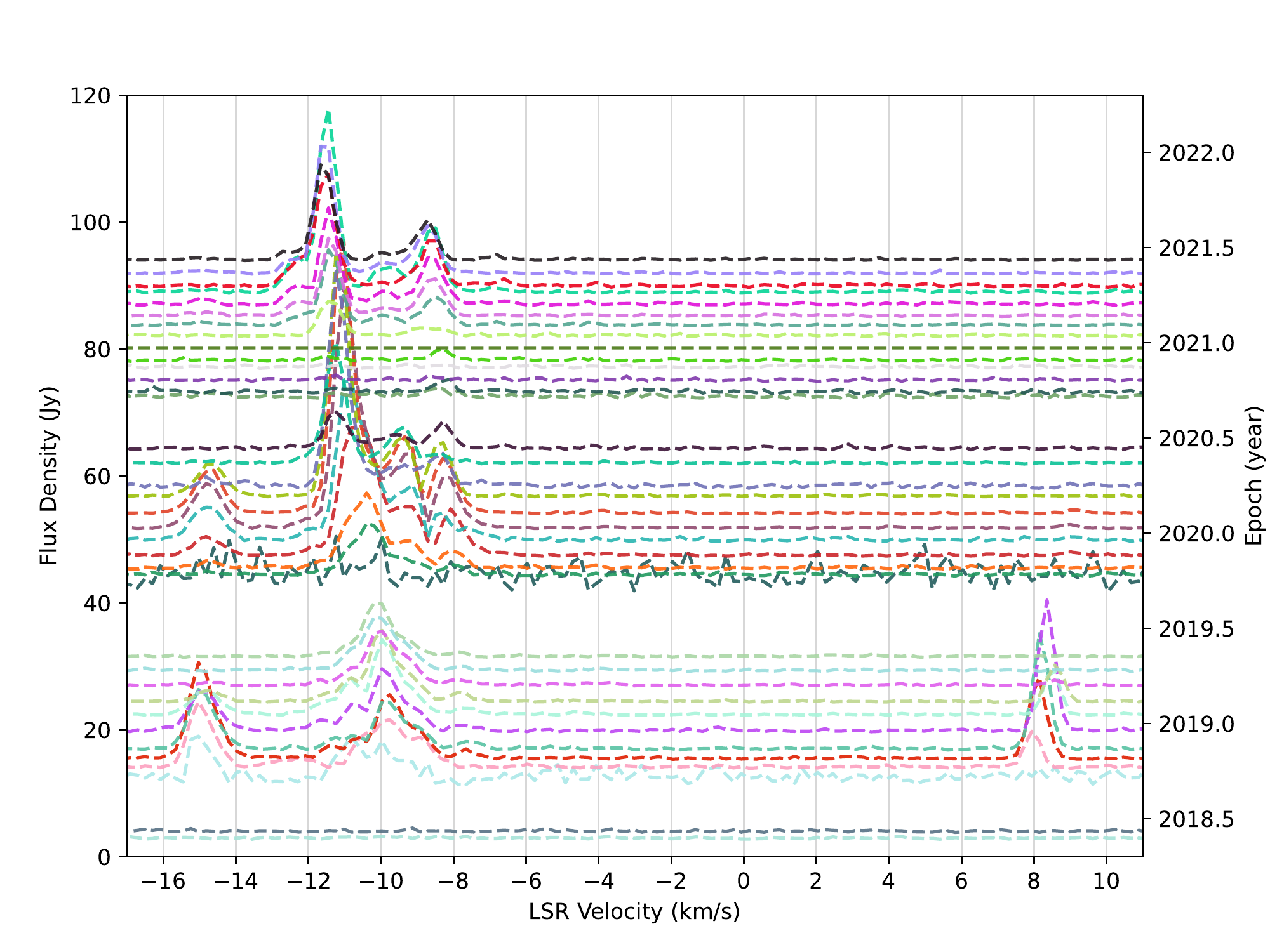}
\caption{The cross power \hho\ maser spectra on the Ky--Ku baseline (305 km). 
Few epochs on Ky--Kt baseline (476 km) when Ku is not available. All of the 37 epochs data were used.
\label{fig:spec}}
\end{figure*}

Frequent observations, as well as the time-span of ESTEMA, allow us to study the  variability and periodicity of maser intensity on timescales of a few weeks to a few years in terms of the possible correlation with stellar pulsation \citep{2010MNRAS.406..395G}.
Figure \ref{fig:spec} demonstrates the changes in the spectrum with time. 
Since some of imaging observations of the masers failed and some of the maser images are made by different array configurations, 
we use the maser intensity on the KVN stations for spectral analysis, e.g.,
making cross-power spectra on a short baseline in Figure \ref{fig:spec}, making the total-power spectra on each KVN stations, and analysing the period and phase of the spectra.

The excitation of \hho\ masers is sensitive to both the dynamics of a stellar wind and the stellar radiation with periodic variation. The previous monitoring observations of \hho\ masers around BX Cam with the VERA Iriki telescope \citep{2008PASJ...60.1077S,2020PASJ...72...56M} did not determine the period and phase lag of the \hho\ maser variability. 

\begin{figure*}[t]
\epsscale{1.15}
\plotone{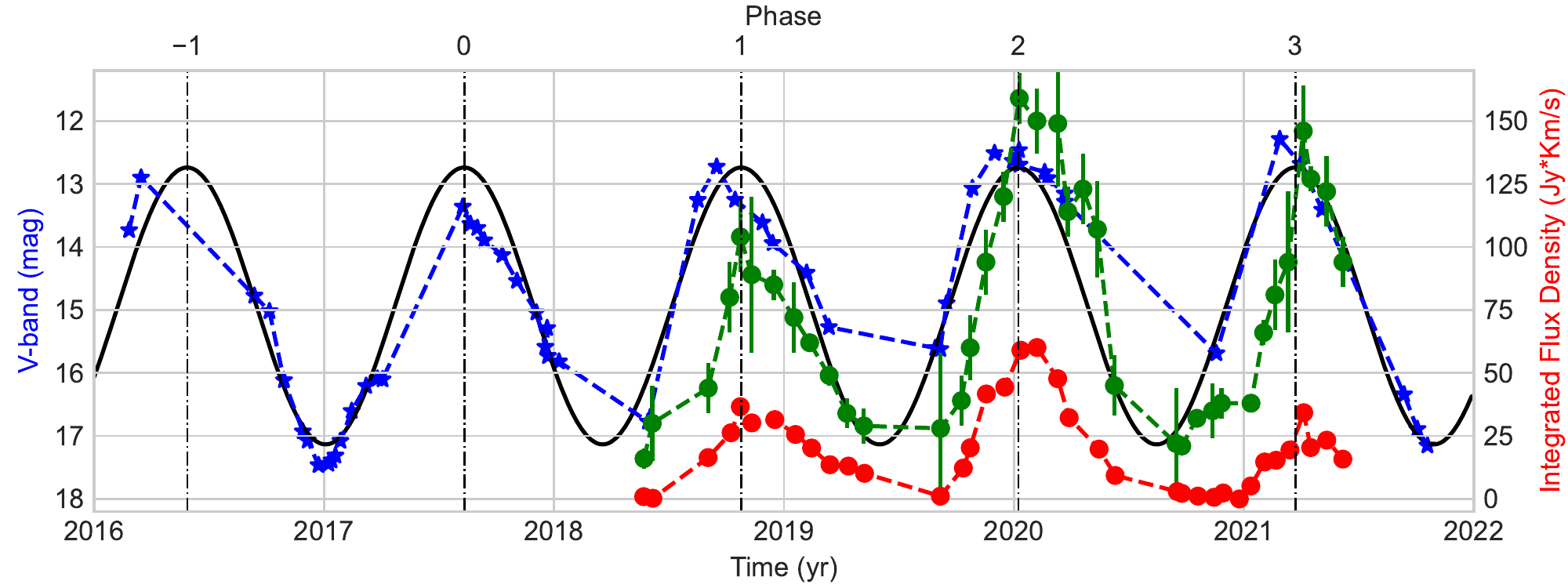}
\caption{
Integrated flux curves of 22\,GHz \hho\ masers from the cross (red circles) and total (green circles) power spectra of the KVN and V-band optical light curve (blue stars) from the AAVSO monitoring data.
The total power spectra is using the weighted average of three KVN antennas (Kt,Ky,Ku), while the cross power spectra is using the short baselines as shown in Figure \ref{fig:spec}.
The \vlsr\ velocity range for estimating the integrated flux densities is from $-$16 to $-$7 \kms and from 7 to 10 \kms.
At K band, the beam size of the KVN single-dish telescope is $\sim$130 as and the synthesized beam size of
the projected KYS-KUS baseline is $\sim$11 mas for BX Cam. 
The black solid line shows the sinusoidal fit to the optical light curve with a period of 440~d. 
\label{fig:curve}}
\end{figure*}

Multiple estimates of the optical period of BX Cam have been reported by the American Association of Variable Star Observers (AAVSO) and \cite{2020PASJ...72...56M}.
AAVSO reported a period of 486~d\footnote{\url{https://www.aavso.org/vsx/index.php?view=detail.top&oid=4634}} without the details of the specific time range of data or a specific method. 
\cite{2020PASJ...72...56M} determined a period of 440~d using the 40 nights in AAVSO database between 2016 February and 2019 December. 
We also reanalysed 54 nights from the AAVSO database at V-band from 2016 February to 2021 October using two methods.
One is the Date Compensated Discrete Fourier Transform (DC DFT) method with the popular analysis tool VStar \citep{2012JAVSO..40..852B}, and another is the Lomb-Scargle Periodogram \citep{1976Ap&SS..39..447L,1982ApJ...263..835S,2018ApJS..236...16V}, which is equivalent to least-squares fitting of sine waves.
We obtained an averaged optical period of 440$\pm$5~d using the two methods as shown in Figure \ref{fig:curve}, which is consistent with \cite{2020PASJ...72...56M}.

During the period from $\phi$ = 0.65 to  3.17 in stellar light curve phase, the \hho\ masers have continuously been found at a local-standard-of-rest velocity (\vlsr) range from $-7$ to $-16$ \kms. We also detected the red-shifted component at \vlsr $=$7--10 \kms\ around the light maxima: from $\phi$ = 1.00 to 1.32  and  from $\phi$ = 1.88 to 2.14, which has also been found in previous observations by \cite{1988A&A...191..283E} and \cite{2020PASJ...72...56M}. 
The peaks of the \hho\ maser intensity appeared at the epoch 2018-10-24 ($\phi$ 1.00), 2020-01-10 ($\phi$ 2.00), and 2021-04-05 ($\phi$ 3.03), 
while the peaks of optical intensity appeared at the 
epoch 2018-09-16 ($\phi$ 0.91), 2020-01-09 ($\phi$ 2.00), and 2021-02-27 ($\phi$ 2.94).
Thus the peaks of the \hho\ maser intensity have an average delayed time-lag  
$\sim$25$\pm9$ days or phase-lag  
$\sim$0.06$\pm0.02$ related to the peaks of optical intensity.
The averaged time interval of \hho\ maser peak intensity is $\sim$447$\pm11$ days, which is a better fit to the period 440~d than 486~d (AAVSO). 

The observed properties of the spectral and flux variability for the \hho\ masers  are as follows.
(1) The \hho\ masers are highly variable so that no maser components were detected in some of the maser flux minimum. (2) The maser flux and the optical magnitude exhibit a similar sinusoidal pattern, showing that the maser flux follows the pulsation of the star \citep[same as the SiO masers][]{2010MNRAS.406..395G}.
(3) The maxima of the periodic maser flux variation do not have a constant magnitude.  
(4) Time lags of the maser flux maxima with respect to the optical light maxima are also expected, but they are not necessarily constant as seen in the data \citep[same as for the SiO masers][]{2010MNRAS.406..395G}. 
(5) The maser flux variations do not follow a specific pattern. The number of spectral features, as well as their fluxes, change not only during a single cycle but also from one cycle to another.
(6) Systematic 
radial velocity drifts
in the maser features are also found as shown on the spectrum.

\subsection{The mapped \texorpdfstring{\hho}{H2O} maser features}

\begin{figure}[t]
\epsscale{1.2}
\plotone{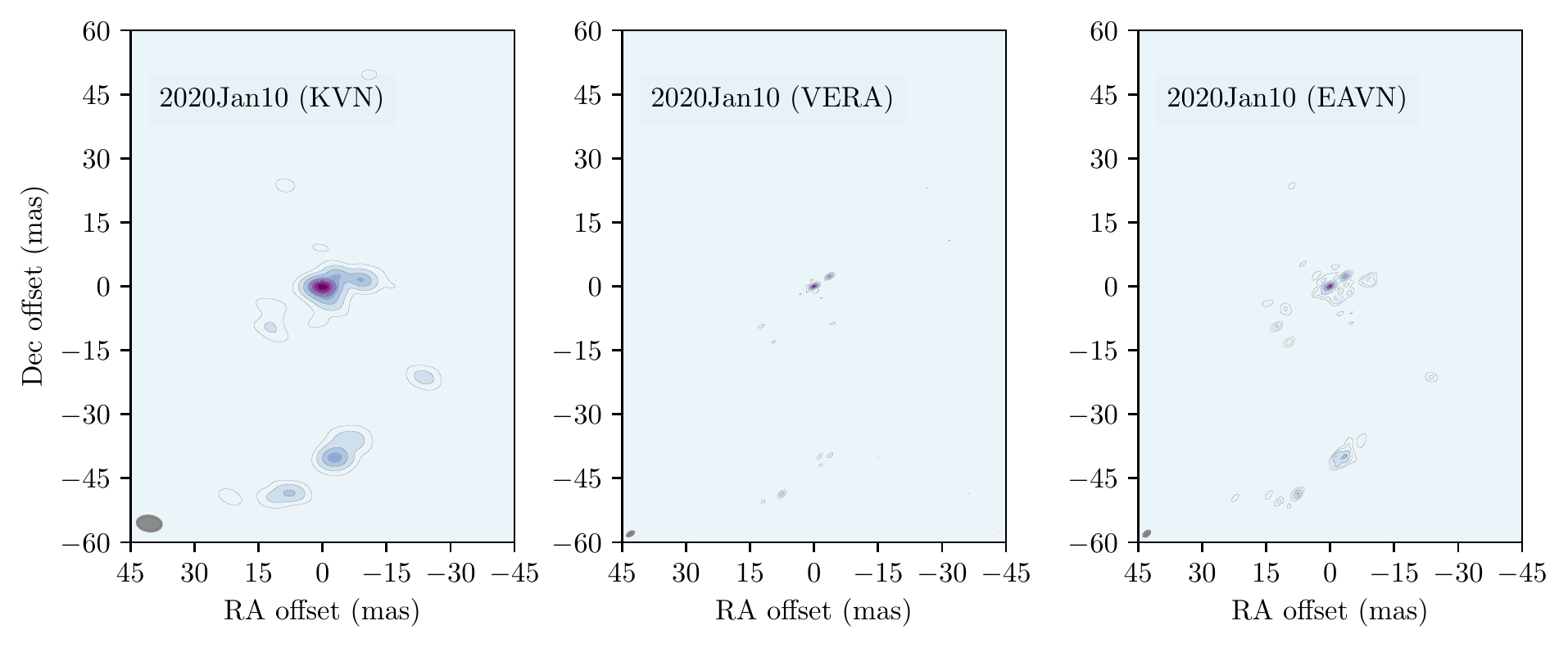}
\caption{Left to Right: Moment-zero maps of the water maser emission in BX Cam from KVN, VERA, and EAVN (KVN+VERA+T6) observations at the same epoch, Jan 10, 2020.
The maps are from images created by collapsing each spectral line cube to a single plane using the AIPS task SQASH, taking the weighted averages of each pixel.
An ellipse in the bottom-left corner of each panel indicates the synthesized beam pattern of the image synthesis. 
The contour map is drawn with contour levels of 3,6,12,24,...×$\sigma$, where 1$\sigma$ corresponds to 0.03 \jybeam (KVN), 0.04 \jybeam (VERA), and 0.03 \jybeam (EAVN), respectively.}
\label{fig:integ}
\end{figure}

Figure \ref{fig:integ} shows the comparison between the maser brightness distributions taken with KVN, VERA and EAVN (KVN+VERA+T6) on 2020Jan10 ($\phi$=2.00). KVN is helpful to detect extended structures, while EAVN can detect a larger number of isolated compact maser features. VERA can measure the bona-fide positional information of bright masers using the dual-beam astrometric system. The \hho\ masers were spatially well resolved into individual maser features with the EAVN synthesized beam. The image including Urumqi baselines shows no obvious difference, since only the brightest and compact maser features 
located at the map origin were detected on $>$ 3000~km long baselines.

The AIPS task SAD was used to extract maser spots information, e.g., positions and brightness, from the EAVN image data cube.
For the component identification, a non-constant Gaussian was used and features with flux higher than six times the noise level were accepted. Maser feature candidates were identified as components if found in more than three consecutive channels within the cube \citep{2010MNRAS.406..395G}. Table \ref{tab:obs} gives the possible  numbers of maser features detected at individual epochs. 
We traced the relative positions and radial velocities of maser features at different epochs, which had been stable from one epoch to another within 1 \kms~in LSR velocity and within 6 \masy $\times \Delta t$ yr in position. The $\Delta t$ is the time gap between two epochs and the limit corresponds to the maximum movement velocity in Table \ref{tab:vel}.
In practice it is difficult to automatically cross-identify components just relying on the above velocity and position window, thus we identified the components manually.

\begin{figure}[t]
\epsscale{1.2}
\plotone{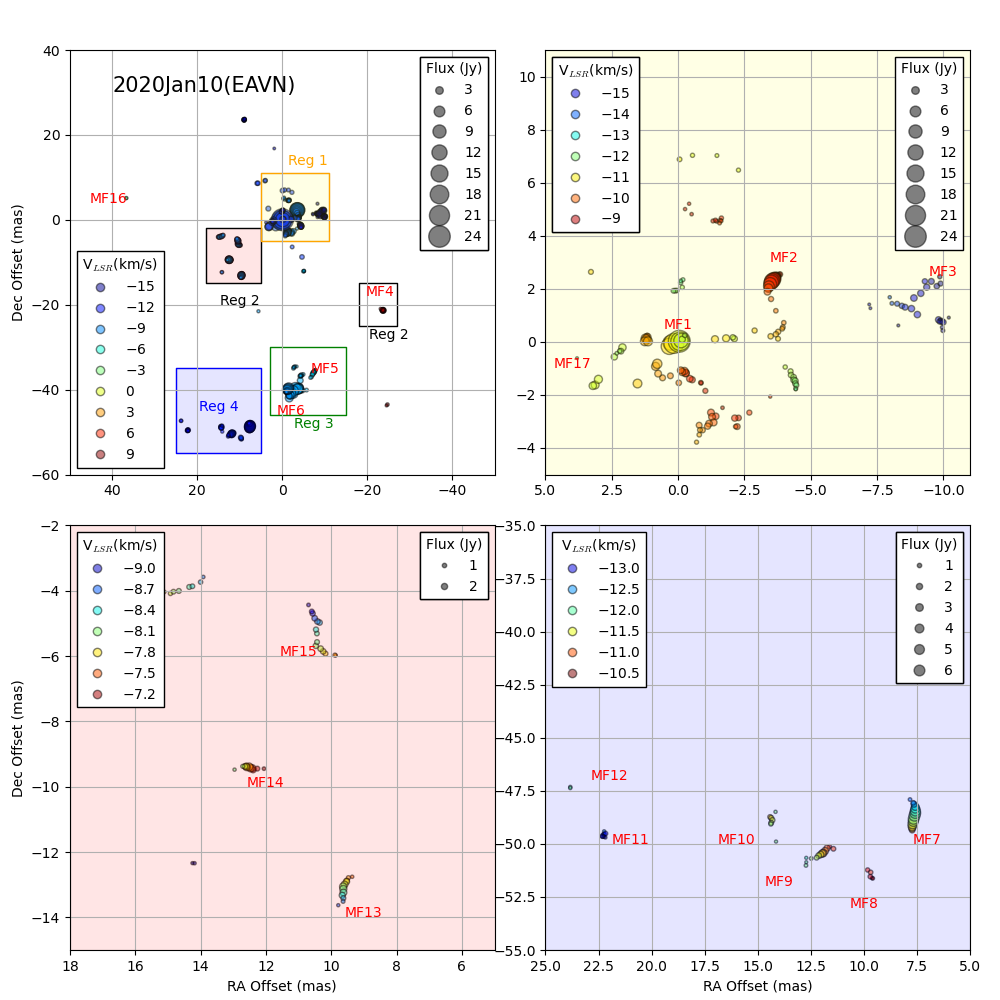}
\caption{EAVN images of the spatial structures of maser features  (top left), with a zoom-in of the colored blocks (top right, bottom left, bottom right), observed on 2020 Jan 10 ($\phi$ 2.00). 
These images are obtained after component fitting from the image data cube with AIPS task SAD.
Each of the filled circles shows a velocity component (maser spot), and each maser feature (MF\#) has a velocity channel spacing of 0.105 \kms.
The position reference feature (MF1) is around the map origin. 
\label{fig:maser}}
\end{figure}

Figure \ref{fig:maser} shows the example of internal structures of the maser features. Among them,  17 maser features with the lifespan longer than one Cycle are labelled with their identifier (MF\#).
Each maser feature is composed of a compact bright part and an extended structure. There are ``spoke-like'' (linearly distributed)  \citep{2012ApJ...744...23Z} maser features at all epochs. These maser features appear to be composed of gas flowing outward from the central star and mostly have decreasing (blue-shifted) or increasing (red-shifted) radial velocities.

\begin{figure}[t]
\epsscale{1}
\plotone{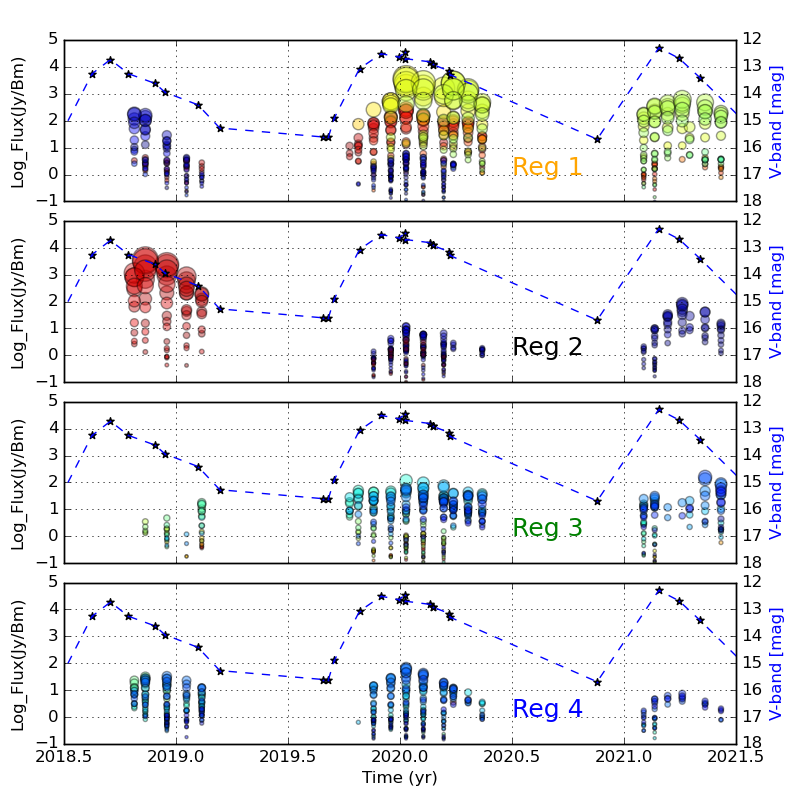}
\caption{Temporal variation of maser features in different regions.
The locations of the different regions are shown in Figure \ref{fig:maser}, while Region 1 includes MF[1,2,3,17], Region 2 includes MF[4,13,14,15], Region 3 includes MF[5,6], Region 4 includes MF[7,8,9,10,11,12]. The velocity gradient is classified by color and the size of maser spots means the flux density. The color and size of maser spots are the same to Fig \ref{fig:timemaser}. Optical light curve found in the AAVSO photometric data (blue stars).
\label{fig:submaser}}
\end{figure}

\begin{deluxetable*}{llrrrrrr}[ht!]
\tablenum{2}
\tablecaption{The 
radial velocity drifts
and relative proper motions of the maser features \label{tab:vel}}
\tablewidth{700pt}
\tabletypesize{\scriptsize}
\tablehead{
\colhead{MF\tablenotemark{a}}  & \colhead{Cycle \tablenotemark{b}}  
& \colhead{Radial Velocity\tablenotemark{c}} & \colhead{RV Drift\tablenotemark{d}} 
& \colhead{RA offset\tablenotemark{c}} & \colhead{Rel \mux \tablenotemark{e}} 
& \colhead{Dec offset\tablenotemark{c}} & \colhead{Rel \muy \tablenotemark{e}} \\
\colhead{ID} & \colhead{Fitting}  
& \colhead{\kms} & \colhead{\kmsyr} 
& \colhead{mas} & \colhead{\masy} 
& \colhead{mas} & \colhead{\masy} }
\startdata
MF1&2+3&-11.25$\pm$0.04&-0.25$\pm$0.05&-0.02$\pm$0.01&0.03$\pm$0.01&0.02$\pm$0.01&-0.02$\pm$0.01\\
     &2&-11.23$\pm$0.06&-0.38$\pm$0.27&-0.02$\pm$0.01&0.04$\pm$0.06&0.03$\pm$0.01&-0.09$\pm$0.04\\
MF2&1+2+3&-9.34$\pm$0.03&-0.35$\pm$0.04&-3.53$\pm$0.01&0.10$\pm$0.02&2.18$\pm$0.02&0.11$\pm$0.03\\
       &2&-9.29$\pm$0.03&-0.70$\pm$0.19&-3.55$\pm$0.01&0.31$\pm$0.07&2.20$\pm$0.02&0.31$\pm$0.12\\
MF3&1+2&-14.78$\pm$0.03&0.17$\pm$0.05&-9.08$\pm$0.05&0.34$\pm$0.09&1.44$\pm$0.04&-0.77$\pm$0.06\\
     &2&-14.79$\pm$0.04&0.31$\pm$0.28&-9.09$\pm$0.08&0.74$\pm$0.55&1.47$\pm$0.06&-1.19$\pm$0.40\\
MF4&1+2&9.07$\pm$0.06&0.79$\pm$0.06&-23.67$\pm$0.16&-1.20$\pm$0.09&-21.26$\pm$0.09&-0.78$\pm$0.10\\
     &2&9.04$\pm$0.04&0.75$\pm$0.35&-23.70$\pm$0.08&-0.49$\pm$0.65&-21.34$\pm$0.04&-0.32$\pm$0.35\\
MF5&1+2+3&-6.95$\pm$0.05&-0.20$\pm$0.09&-7.46$\pm$0.11&-0.72$\pm$0.19&-35.87$\pm$0.12&-2.25$\pm$0.21\\
       &2&-6.89$\pm$0.11&-0.33$\pm$0.77&-7.48$\pm$0.23&-1.39$\pm$1.56&-35.85$\pm$0.24&-3.90$\pm$1.64\\
MF6&1+2+3&-8.32$\pm$0.02&-0.32$\pm$0.04&-2.75$\pm$0.08&-0.48$\pm$0.13&-39.83$\pm$0.09&-3.75$\pm$0.15\\
        &2&-8.32$\pm$0.03&-0.73$\pm$0.17&-2.67$\pm$0.12&0.34$\pm$0.61&-39.94$\pm$0.13&-4.30$\pm$0.70\\
MF7&1+2+3&-11.87$\pm$0.03&-0.38$\pm$0.06&7.60$\pm$0.01&0.64$\pm$0.01&-48.59$\pm$0.04&-3.08$\pm$0.07\\
       &2&-11.84$\pm$0.04&-0.68$\pm$0.24&7.62$\pm$0.01&0.47$\pm$0.05&-48.63$\pm$0.05&-3.25$\pm$0.27\\
MF8&1+2+3&-10.86$\pm$0.04&-0.27$\pm$0.04&9.77$\pm$0.03&1.50$\pm$0.03&-51.22$\pm$0.03&-4.82$\pm$0.03\\
       &2&-10.77$\pm$0.08&-0.54$\pm$0.40&9.75$\pm$0.10&1.75$\pm$0.47&-51.34$\pm$0.09&-4.97$\pm$0.44\\
MF9&1+2+3&-11.31$\pm$0.03&0.01$\pm$0.04&11.91$\pm$0.03&1.24$\pm$0.05&-50.25$\pm$0.02&-3.65$\pm$0.04\\
       &2&-11.34$\pm$0.04&0.20$\pm$0.21&11.95$\pm$0.04&1.13$\pm$0.19&-50.37$\pm$0.03&-3.41$\pm$0.16\\
MF10&1+2&-11.58$\pm$0.03&-0.31$\pm$0.05&14.25$\pm$0.03&1.30$\pm$0.05&-48.84$\pm$0.03&-3.46$\pm$0.06\\
      &2&-11.56$\pm$0.04&-0.48$\pm$0.26&14.27$\pm$0.05&1.10$\pm$0.27&-48.84$\pm$0.05&-3.48$\pm$0.30\\
MF11&1+2&-12.86$\pm$0.04&-0.31$\pm$0.06&22.18$\pm$0.04&2.78$\pm$0.05&-49.47$\pm$0.05&-4.26$\pm$0.07\\
      &2&-12.85$\pm$0.04&-0.50$\pm$0.32&22.21$\pm$0.02&2.21$\pm$0.16&-49.52$\pm$0.02&-3.29$\pm$0.17\\
MF12&1+2&-11.87$\pm$0.32&0.00$\pm$0.31&23.88$\pm$0.12&3.29$\pm$0.12&-47.21$\pm$0.16&-4.34$\pm$0.15\\
      &1&-11.87$\pm$0.83&0.00$\pm$0.80&23.45$\pm$0.31&2.88$\pm$0.29&-47.08$\pm$0.40&-4.21$\pm$0.39\\
MF13&2+3&-8.20$\pm$0.04&-0.49$\pm$0.05&9.57$\pm$0.03&1.31$\pm$0.03&-13.12$\pm$0.03&-0.66$\pm$0.04\\
      &2&-8.13$\pm$0.05&-1.01$\pm$0.26&9.53$\pm$0.02&1.66$\pm$0.10&-13.07$\pm$0.05&-1.01$\pm$0.25\\
MF14&2+3&-7.61$\pm$0.03&-0.89$\pm$0.08&12.49$\pm$0.02&1.78$\pm$0.06&-9.46$\pm$0.02&-0.80$\pm$0.05\\
      &2&-7.58$\pm$0.03&-1.29$\pm$0.19&12.50$\pm$0.02&1.64$\pm$0.13&-9.49$\pm$0.02&-0.52$\pm$0.11\\
MF15&2+3&-8.16$\pm$0.05&-0.59$\pm$0.11&10.40$\pm$0.03&1.30$\pm$0.07&-5.55$\pm$0.05&-0.45$\pm$0.12\\
       &2&-8.15$\pm$0.05&-0.83$\pm$0.37&10.42$\pm$0.03&1.00$\pm$0.23&-5.57$\pm$0.05&0.10$\pm$0.39\\
MF16&2+3&-3.98$\pm$0.05&-0.23$\pm$0.08&36.68$\pm$0.03&3.04$\pm$0.05&5.14$\pm$0.03&0.75$\pm$0.05\\
      &2&-3.93$\pm$0.09&-0.46$\pm$0.35&36.60$\pm$0.06&3.37$\pm$0.22&5.15$\pm$0.08&0.74$\pm$0.31\\
MF17&1+2+3&-11.41$\pm$0.03&-0.52$\pm$0.05&2.96$\pm$0.04&0.73$\pm$0.05&-1.48$\pm$0.05&-0.05$\pm$0.07\\
        &2&-11.35$\pm$0.05&-0.82$\pm$0.25&2.97$\pm$0.06&0.51$\pm$0.28&-1.41$\pm$0.08&-0.53$\pm$0.36\\
\enddata
\tablenotetext{a}{Maser feature ID from Figure \ref{fig:maser}.}
\tablenotetext{b}{Each Cycle corresponds to the epochs:  Cycle 1 ($\phi\ 1.00\sim 1.25$) , Cycle 2 ($\phi\ 1.89\sim 2.29$), and Cycle 3 ($\phi\ 2.89\sim 3.17$). }
\tablenotetext{c}{The reference epoch is J2020.0 and the reference position is MF1. The fitted positions of MF1 with minor offsets are caused by the maser structures.} 
\tablenotetext{d}{The radial velocity drift} 
\tablenotetext{e}{The relative proper motions are defined as $\mux = \mu_{\alpha \cos{\delta}}$ and $\muy = \mu_{\delta}$.} 
\end{deluxetable*}

\subsection{Astrometry}
\label{sec:metry}

\subsubsection{VERA Bona-Fide Astrometry in ESTEMA 
\label{sec:plx}}
\begin{figure}[t]
\epsscale{0.5}
\plotone{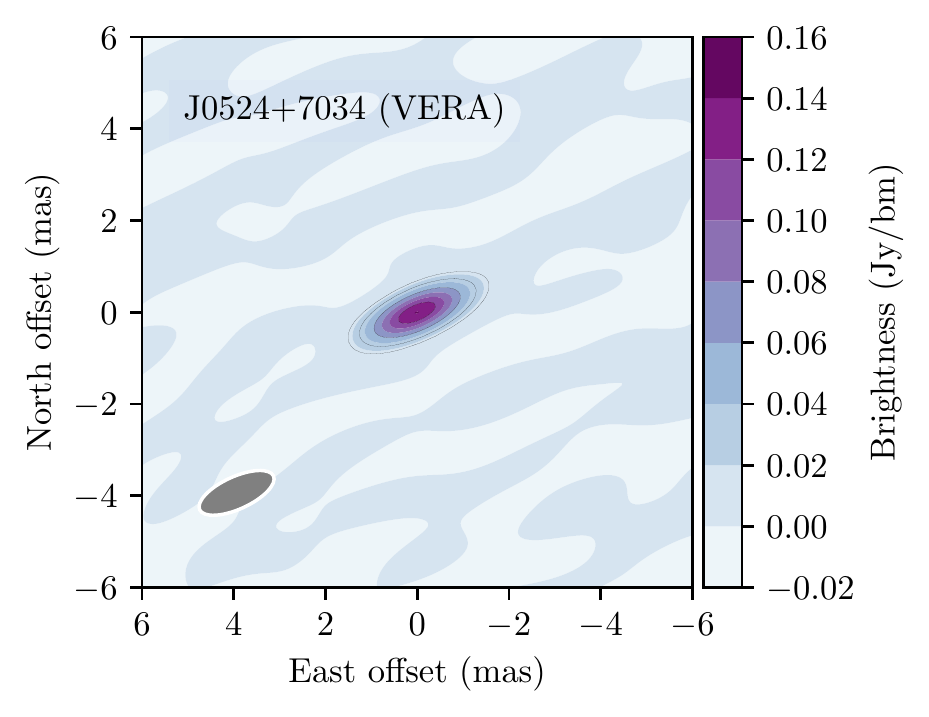}
\caption{VERA image of J0524$+$7034 observed on 2020 January 10. The contour map is drawn with contour levels of 3,6,12,24,...×$\sigma$, where 1$\sigma$ corresponds to 0.005 \jybeam.
\label{fig:qso}}
\end{figure}

As marked in Table \ref{tab:obs},  
14 epochs of data yielded successful phase-referencing astrometry with VERA dual-beam system.
The dual-beam astrometry failed at other epochs due to various reasons, 
such as the masers in the trough of the stellar pulsation cycles being too weak, key stations missing, and low sensitivity due to bad weather. The compact calibrator J0524$+$7034 shown in Figure~\ref{fig:qso} was used as the phase reference source, thus we expect no or little effects from its source structure.

As shown in Figure \ref{fig:integ}, few maser features can be detected with VERA alone. 
Among them, MF1 is the brightest maser feature, which can be  detected over one year. 
The maser feature position can not be well determined using only one maser spot, because the maser spots are blended in position and frequency and the intensity of the blended components can vary with time (as shown in Figure \ref{fig:timemaser}). Instead, we used positions in about 5 spectral channels  (velocity spacing of $\sim$0.5\kms) where the maser spots were at the peak brightness, with compact emission and minimal blending.  The positions of the maser spots relative to the calibrator J0524$+$7034 were measured  as a function of time, and the positions were then modeled by the parallax, proper motion, position at a reference epoch, and the possible acceleration. We then assess the astrometric quality based on the magnitude of the post-fit residuals. The formal position uncertainties were usually smaller than the actual error, since the unknown systematic error (i.e.,  the tropospheric delay residual, maser structures, etc.) often dominate over random noise \citep{2009ApJ...693..397R,2020PASJ...72...52N}. The systematic error was added in quadrature to the priori error so that the reduced chi-square value became nearly unity.

\begin{figure*}[t]
\epsscale{1}\plotone{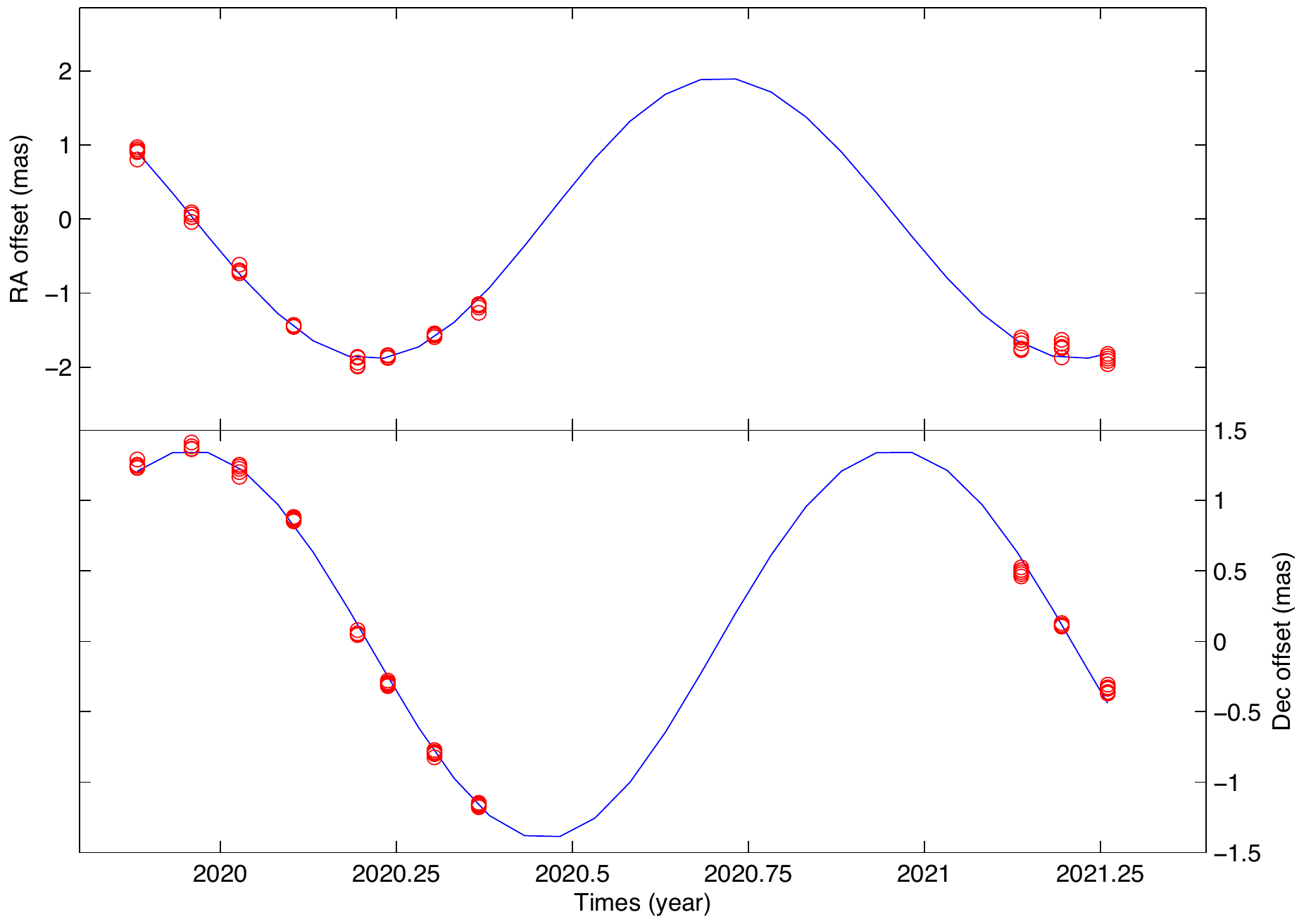}
\epsscale{0.55}\plotone{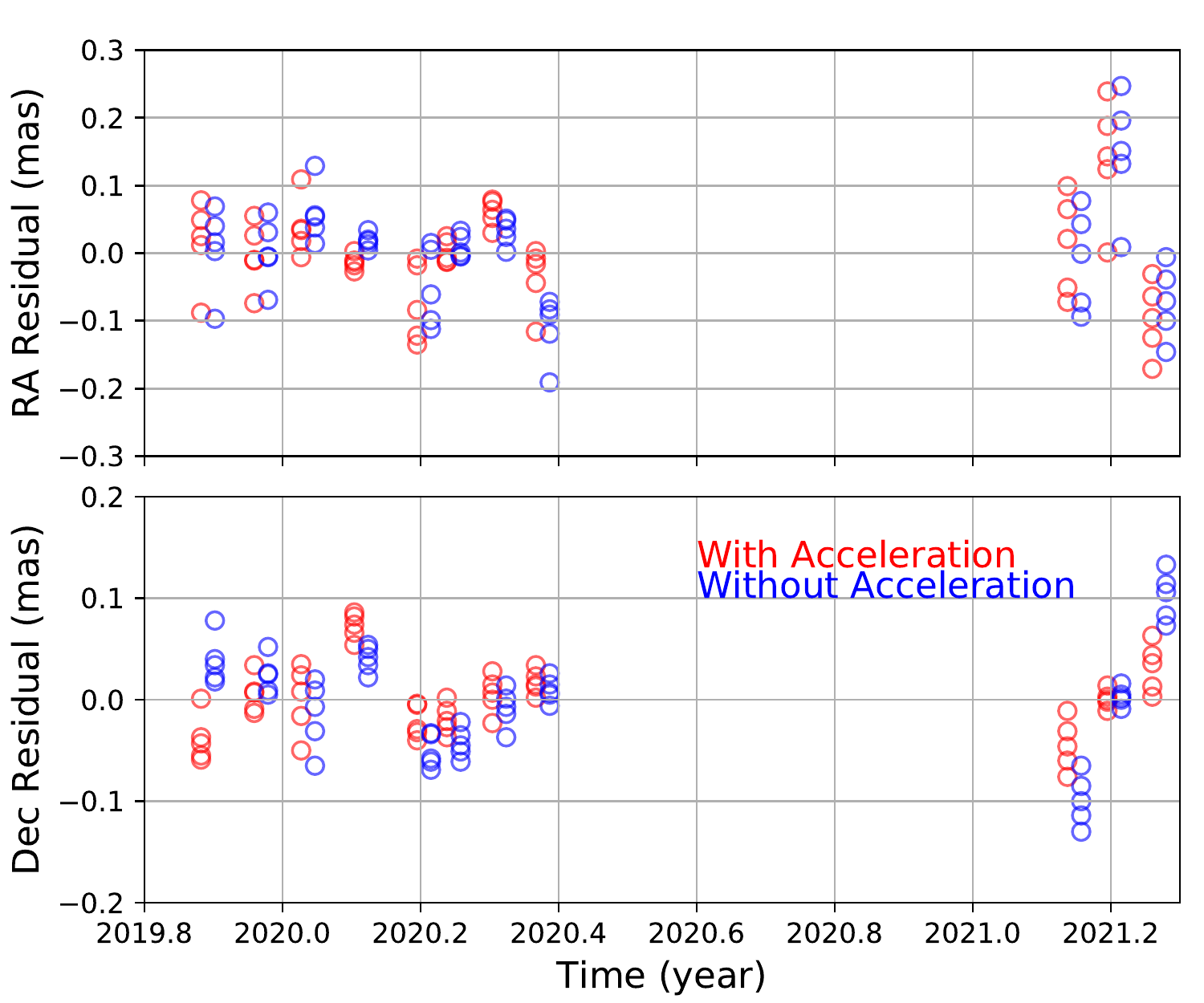}
\epsscale{0.55}\plotone{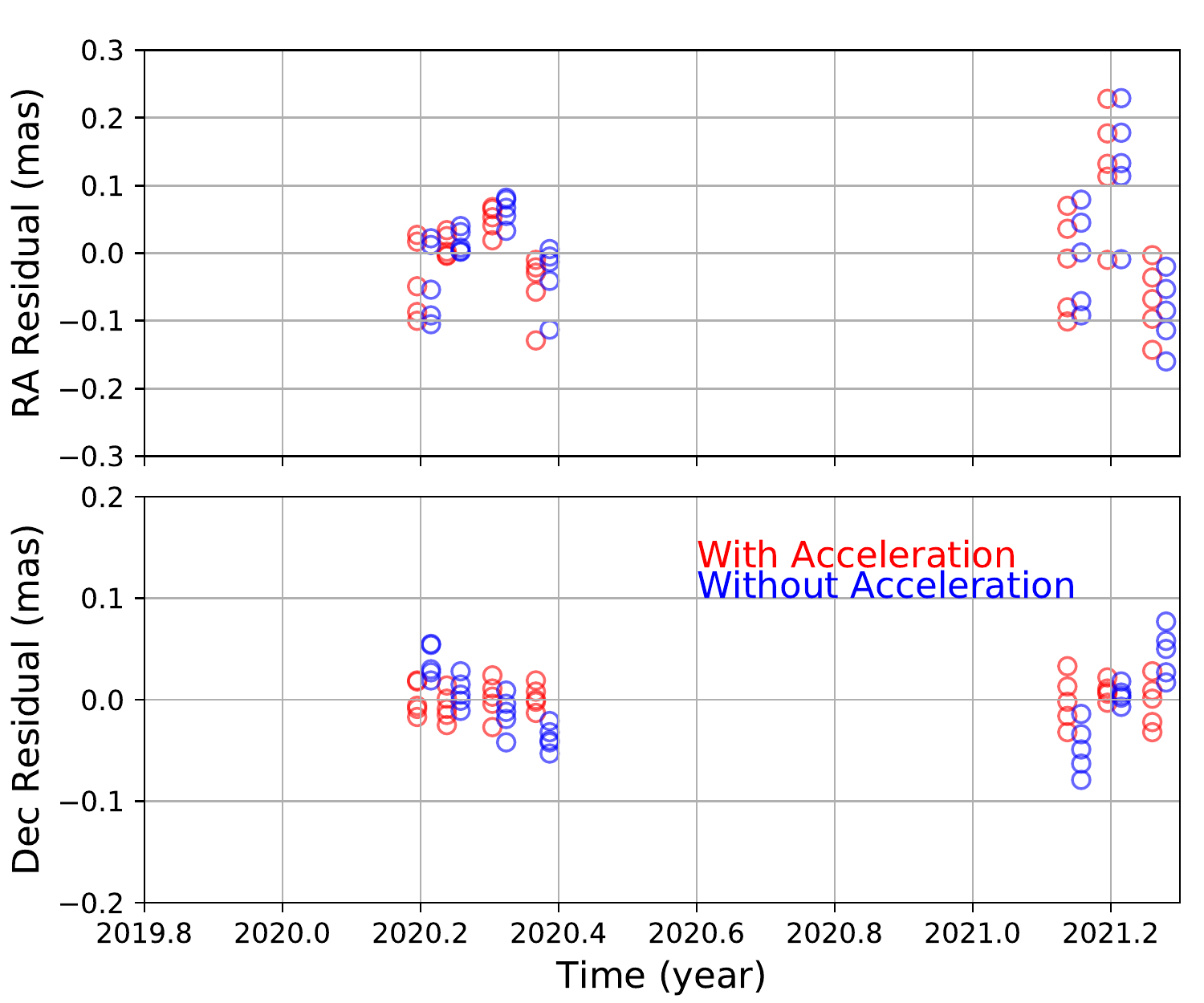}
\caption{Parallax fitting for MF1 with VERA astrometry. 
Top panel: An example of the fitted parallax curve using the data points during the period 2019.8--2021.3 without an acceleration. 
Bottom-left panel: Residuals of the fitting using the data points during the period 2019.8--2021.3. The data points from the fitting including an acceleration have a constant horizontal offset from those without an acceleration for clarity. Bottom-right panel: Same as the bottom-left panel but the period 2020.2--2021.3.  
\label{fig:plxmf1}}
\end{figure*}

Figure \ref{fig:plxmf1} and Table \ref{tab:mf1_fit} show the outcomes of the astrometric fitting for MF1 positions measured with the VERA astrometric technique \citep{2020PASJ...72...52N}. 
In this astrometric fitting, we compared the different parameters, such as, using the data at different timescales and including the acceleration or not.
We found different deviations from a common constant proper motion, 
or an acceleration, in declination between 2020 and 2021. This could come from either the variation in the maser structure (Figure \ref{figmf}) or a non-uniform acceleration in the maser feature. 
Taking into account the ambiguity in the results from these different estimation procedures, 
we conservatively obtained a parallax value of 1.79$\pm$0.08 mas. This is consistent with 1.76$\pm$0.10 mas from \Gaia\ EDR3 \citep{2021A&A...649A...1G} and  1.73$\pm$0.03 mas from the previous VERA result   \citep{2020PASJ...72...56M} within 1 sigma level.
The proper motions of the individual maser features
usually are different from the central star, due to the extra internal motions of the masers.
However, the proper motion of the maser feature MF1 (14.37$\pm$0.20, -35.46$\pm$ 0.44 \masy) is consistent with \Gaia\ EDR3 (14.29$\pm$0.07, -34.63$\pm$ 0.10 \masy) within 1 sigma in east direction  and 2 sigma in north direction, due to a small projection distance (less than 7 mas as discussed in Section \ref{sec:sfpr}) between MF1 and central star.

\begin{deluxetable*}{llllrrrrr}[ht!]
\tablenum{3}
\tablecaption{Astrometric fitting for MF1 with VERA  \label{tab:mf1_fit}}
\tablewidth{700pt}
\tabletypesize{\scriptsize}
\tablehead{
\colhead{MF\tablenotemark{a}}  & \colhead{Time} & \colhead{Acceleration}& 
\colhead{Parallax } & \colhead{\mux \tablenotemark{b}} & 
\colhead{\muy \tablenotemark{b}} & \colhead{Acc x\tablenotemark{c}} & 
\colhead{Acc y\tablenotemark{c}} & \colhead{RMS RA/Dec \tablenotemark{d}}  \\ 
\colhead{ID} & \colhead{Range}& \colhead{Estimation} & \colhead{(mas)}  & 
\colhead{mas~yr$^{-1}$} & \colhead{mas~yr$^{-1}$}& \colhead{mas~yr$^{-2}$} & \colhead{mas~yr$^{-2}$} &  \colhead{(mas  mas)}
} 
\startdata
   MF1  & 2019.8-2021.3 & No  &   1.90$\pm$0.02 &  14.35$\pm$0.02  &  $-35.13\pm$0.02 & - & - & 0.082 0.060 \\ 
        & 2019.8-2021.3 & Yes &   1.82$\pm$0.03 &  14.08$\pm$0.06  &  $-35.51\pm$0.10 & 0.47$\pm$0.10 &0.70$\pm$0.19  &  0.081 0.040\\
        & 2020.2-2021.3& No   &   1.75$\pm$0.05 &  14.40$\pm$0.04  &  $-35.04\pm$0.03 & - & - &  0.093 0.043 \\ 
        & 2020.2-2021.3& Yes  &   1.67$\pm$0.07 &  14.65$\pm$0.25  &  $-36.15\pm$0.23 & -0.36$\pm$0.34 &1.53$\pm$0.32  & 0.093 0.027 \\ 
\hline        
  & Combined\tablenotemark{e}    &     &   1.79$\pm$0.08    & 14.37$\pm$0.20 & -35.46$\pm$0.44\\
\enddata
\tablenotetext{a}{Maser feature ID from Figure \ref{fig:maser}.}
\tablenotetext{b}{Absolute proper motions are defined as $\mux = \mu_{\alpha \cos{\delta}}$ and $\muy = \mu_{\delta}$.} 
\tablenotetext{c}{The fitted accelerations/decelerations.} 
\tablenotetext{d}{The root mean square (RMS) of the residuals in RA and Dec.} 
\tablenotetext{e}{The unweighted average and standard deviation.} 
\end{deluxetable*}

\subsubsection{Astrometric fitting for maser features with the EAVN data}
\label{sec:plx-other}

We can obtain the absolute positions of the bright and compact maser spots from the 14 epochs with the VERA data. 
The bright and compact maser feature (MF1 in $\phi\ 1.89\sim 2.29$ and $\phi\ 2.93\sim 3.03$ , MF4 in $\phi\ 1.04\sim 1.19$) was then used to register the  VERA and EAVN maps. 
The registration accuracy can be expressed as $\sigma_{registration} = \sqrt{ {\sigma_{{thermal_{\rm VERA}}}}^2 + {\sigma_{{structure_{\rm VERA}}}}^2 + {\sigma_{{thermal_{\rm EAVN}}}}^2 + {\sigma_{{structure_{\rm EAVN}}}}^2}$
, where $\sigma_{thermal}$ is the thermal (random) noise
determined by an elliptical Gaussian fitting with the AIPS task ``SAD''
and $\sigma_{structure}$ is the error effected by maser structure estimated by the blended peak components as shown in Figure \ref{fig:timemaser} in the VERA/EAVN images.  Because MF1 and MF4 are bright and compact, the thermal noises are $<$ 0.02 mas and structure errors are $\sim$ 0.05 mas for both VERA and EAVN maps. We believe the registration errors between VERA and EAVN maps are typically $\sim$ 0.1 mas. 
These registration errors are used as the prior errors in the astrometric fitting.
Note that this uncertainty  may be propagated to the errors in deriving the parallaxes of maser features with the astrometrically registered EAVN maps, as described in Section \ref{subsection4.1-astrometry}.

Table~\ref{tab:fitall} shows the independent parallax fits for the residuals of maser feature positions derived after subtracting the modulation of the annual parallax derived from the VERA data.
The results  
were based on the same fitting procedures as mentioned in Section \ref{sec:plx} and with/without acceleration parameters were compared. Averaging all the fitting results, we obtained a combined parallax residual of 0.07$\pm$0.06 mas, which is consistent with the estimated error of the parallax ($\pi=$ 1.79$\pm$0.08 mas). 

\begin{deluxetable*}{lllrrrrrrr}[ht!]
\tablenum{4}
\tablecaption{Astrometric fitting for positions of the maser features with EAVN, registered using VERA astrometry \label{tab:fitall}}
\tablewidth{700pt}
\tabletypesize{\scriptsize}
\tablehead{
\colhead{MF\tablenotemark{a}}  & \colhead{Time} & \colhead{Acceleration} & 
\colhead{Residual \tablenotemark{b}}  & \colhead{\mux \tablenotemark{c}} & 
\colhead{\muy \tablenotemark{c}} & \colhead{Acc x\tablenotemark{d}} & 
\colhead{Acc y\tablenotemark{d}} & \colhead{RMS RA/Dec \tablenotemark{e}}  \\ 
\colhead{ID} &   \colhead{Range (Epochs)} & \colhead{Estimation} & \colhead{Parallax (mas)} & 
\colhead{mas~yr$^{-1}$} & \colhead{mas~yr$^{-1}$}& \colhead{mas~yr$^{-2}$} & \colhead{mas~yr$^{-2}$}  &  \colhead{(mas  mas)}
} 
\startdata
MF1  & 2019.8-2021.3 (11) & Yes    &  0.00$\pm$0.04  & 14.08$\pm$0.06  &  -35.51$\pm$0.10 & 0.47$\pm$0.10 &0.70$\pm$0.19  &  0.081 0.040\\   
                          & &  No  &  0.00$\pm$0.03  & 14.35$\pm$0.02  &  -35.13$\pm$0.02 & - & - & 0.082 0.060 \\ 
MF2  & 2018.8-2021.3 (13)  &  Yes  &  0.13$\pm$0.04  & 14.40$\pm$0.02  &  -34.92$\pm$0.02 & 0.14$\pm$0.05 &-0.48$\pm$0.05  & 0.075 0.094 \\ 
                          & &  No  &  0.05$\pm$0.03  & 14.38$\pm$0.02  &  -34.85$\pm$0.03 & - & - & 0.083 0.161 \\ 
MF3  & 2018.8-2021.2 (11)   &  Yes  & 0.06$\pm$0.09  & 14.79$\pm$0.51 &  -36.43$\pm$0.23 & -0.28$\pm$0.28 & -1.82$\pm$0.45  & 0.601 0.470\\
                          & &  No  &  0.11$\pm$0.06  & 14.62$\pm$0.08 &  -35.51$\pm$0.05 &   - & - & 0.609 0.446 \\
MF4  & 2018.8-2020.2 (8)   &  Yes  &  0.09$\pm$0.05  & 13.46$\pm$0.18  &  -35.77$\pm$0.20 & 0.73$\pm$0.26 & 0.09$\pm$0.32 & 0.059 0.104 \\  
                          & &  No  &  0.02$\pm$0.04  & 13.01$\pm$0.03  &  -35.84$\pm$0.07 & - & - & 0.057 0.103 \\ 
MF5  & 2019.0-2021.2 (5)    &  Yes  & 0.07$\pm$0.11  & 13.56$\pm$0.12 &  -37.50$\pm$0.17 &  0.66$\pm$0.26 &  0.29$\pm$0.34 & 0.314 0.362\\
                          & &  No  &  0.02$\pm$0.09  & 13.85$\pm$0.17 &  -37.34$\pm$0.11 &   - & - &  0.354 0.360\\
MF7 & 2018.9-2021.3 (13)   &  Yes  &  0.17$\pm$0.07  & 14.81$\pm$0.17  &  -38.22$\pm$0.22 & 0.35$\pm$0.29 &-0.62$\pm$0.40  & 0.078 0.189 \\  
                          & &  No  &  0.09$\pm$0.04  & 14.99$\pm$0.04  &  -38.51$\pm$0.05 & - & - & 0.082 0.192 \\ 
MF8  & 2018.8-2021.2 (9)  &  Yes    & 0.12$\pm$0.09  & 15.78$\pm$0.05 &  -39.65$\pm$0.04 &  0.14$\pm$0.11 & -0.04$\pm$0.14 & 0.179 0.141 \\
                          & &  No  &  0.08$\pm$0.05  & 15.78$\pm$0.05 &  -39.64$\pm$0.03 &  - & - &  0.180 0.141\\
MF9  & 2018.8-2021.2 (12)  &  Yes   & 0.05$\pm$0.06  & 15.32$\pm$0.05 &  -38.70$\pm$0.05 & -0.17$\pm$0.09 &  0.03$\pm$0.11 & 0.294 0.172\\
                          & &  No  & -0.01$\pm$0.04  & 15.37$\pm$0.04 &  -38.70$\pm$0.03 &  - & - & 0.298 0.172 \\
MF10 & 2019.8-2021.3 (10)  &  Yes   & 0.17$\pm$0.09  & 15.65$\pm$0.26 &  -38.28$\pm$0.24 &  0.16$\pm$0.47 & -0.58$\pm$0.48 & 0.139 0.198 \\
                          &  &  No &  0.11$\pm$0.06  & 15.59$\pm$0.07 &  -38.02$\pm$0.08 &  - & - &  0.139 0.200 \\
MF11 & 2018.8-2021.3 (7)  &  Yes    & 0.14$\pm$0.09  & 16.59$\pm$0.29 &  -38.87$\pm$0.27 & -0.97$\pm$0.65 &  0.74$\pm$0.52 & 0.115 0.116\\
                          &  & No &   0.11$\pm$0.05  & 17.06$\pm$0.05 &  -39.21$\pm$0.07 &   - &  - & 0.127 0.129\\
MF12 & 2018.8-2020.2 (4)  &  Yes  &   0.03$\pm$0.05  & 17.96$\pm$0.37 &  -38.95$\pm$0.48 &  0.86$\pm$0.47 & -0.15$\pm$0.62 & 0.120 0.182\\
                          &  &  No  & 0.04$\pm$0.05  & 17.48$\pm$0.18 &  -39.00$\pm$0.10 &   - & - & 0.121 0.184 \\
MF13 & 2018.8-2020.0 (10)  &  Yes  &  0.05$\pm$0.04  & 15.75$\pm$0.07  &  -36.26$\pm$0.15 & -0.32$\pm$0.12 &0.90$\pm$0.28  & 0.169 0.228 \\ 
                          &  &  No  & 0.01$\pm$0.03  & 15.60$\pm$0.01  &  -35.76$\pm$0.02 & - & - & 0.169 0.235 \\ 
MF14 & 2019.9-2021.2 (9)  &  Yes  &   0.20$\pm$0.07  & 15.86$\pm$0.21 &  -35.23$\pm$0.28 &  0.52$\pm$0.37 & -1.35$\pm$0.50 & 0.169 0.101 \\
                          &  &  No  & 0.05$\pm$0.04  & 16.04$\pm$0.08 &  -35.95$\pm$0.08 &  - & - & 0.184 0.130 \\
MF15 & 2019.8-2021.2 (7)  &  Yes  &   0.11$\pm$0.12  & 14.90$\pm$0.42 &  -34.94$\pm$0.32 &  1.28$\pm$0.65 & -1.21$\pm$0.55 & 0.208 0.308\\
                          &  &  No  & 0.13$\pm$0.06  & 15.71$\pm$0.08 &  -35.60$\pm$0.10 &   - & - & 0.210 0.314 \\
MF16 & 2020.1-2021.2 (4)  &  Yes  &   0.07$\pm$0.14  & 16.76$\pm$0.41 &  -34.57$\pm$0.46 &  1.03$\pm$0.60 &  0.19$\pm$0.81 & 0.070 0.070\\ 
                          &  &  No  & 0.00$\pm$0.09  & 17.46$\pm$0.10 &  -34.44$\pm$0.11 &  - & - &  0.070 0.068\\
MF17 & 2019.0-2021.3 (12)  &  Yes  &  0.06$\pm$0.07  & 14.75$\pm$0.08 &  -35.16$\pm$0.11 &  0.46$\pm$0.14 & -0.09$\pm$0.19 & 0.233 0.322\\
                          &  &  No  &-0.02$\pm$0.05  & 15.03$\pm$0.05 &  -35.18$\pm$0.07 &   - & - & 0.250 0.321 \\
\hline
  & Combined\tablenotemark{f}    &     Yes &   0.10$\pm$0.05 \\
  & Combined\tablenotemark{f}    &     No  & 0.05$\pm$0.05\\
  & Combined\tablenotemark{f}    &     ALL &   0.07$\pm$0.06\\
\enddata
\tablenotetext{a}{Maser feature ID from Figure \ref{fig:maser}. }
\tablenotetext{b}{The residual parallax after subtracting the parallax 1.82$\pm$0.03 mas for with acceleration(Yes) and 1.90$\pm$0.02 mas for without acceleration(No).}
\tablenotetext{c}{The relative proper motions are defined as $\mux = \mu_{\alpha \cos{\delta}}$ and $\muy = \mu_{\delta}$.} 
\tablenotetext{d}{The possible accelerations/decelerations.} 
\tablenotetext{e}{The root mean square (RMS) of the residuals in RA and Dec.} 
\tablenotetext{f}{The unweighted average and standard deviation.}
\end{deluxetable*}

\subsubsection{The stellar position in the \texorpdfstring{\hho}{H2O} maser distribution}
\label{sec:sfpr}

\begin{figure}[t]
\epsscale{0.7}
\plotone{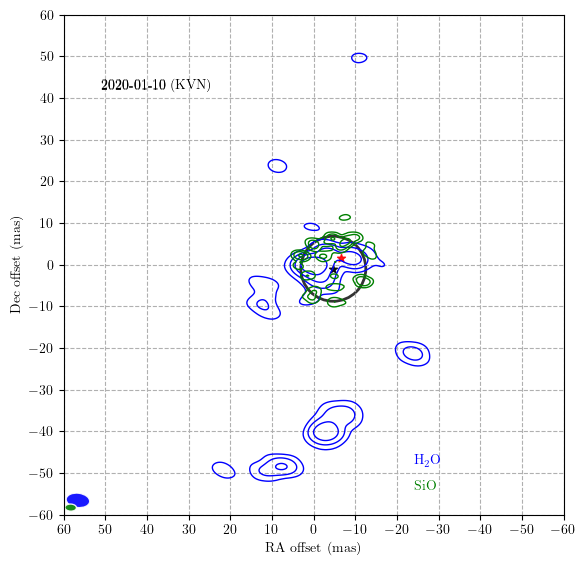}
\caption{Registered map of the \hho\ and SiO({\it$v$=1, J=1$\rightarrow$0}) masers observed on 2020 January 10 using the KVN data, which were taken from velocity-channel
integration with weights based on the root-mean-square noise level.
Black star indicates  the estimated position of the star at the center of the SiO maser ring
using the least-squares circle method \citep{2018NatCo...9.2534Y}.
 Red star indicates the position of the star derived from \Gaia\ EDR3. 
The contour map is drawn with contour levels of 3,6,12,24,...×$\sigma$, where 1$\sigma$ corresponds to 0.025 \jybeam (\hho) and 0.120 \jybeam (SiO), respectively.
\label{fig:sio}}
\end{figure}

Determining the stellar position with respect to the \hho\ maser distribution is important to estimate the center of 3-D  kinematics. 
\Gaia\ EDR3 provided the stellar position with optical astrometry, however there may exist uncertainty for this AGB star \citep{2019ApJ...875..114X}.
On the other hand, the center of the SiO maser ring should correspond to the position of the star.  Figure \ref{fig:sio} shows the registered map of the \hho\ and SiO masers with the KVN data, to which the SFPR calibration technique was applied \citep{2014AJ....148...97D}. The positional reference is the \hho\ maser at $V_{\rm LSR} = -$11.05 \kms\ (MF1). We obtained the position of the central star ($-5.1\pm 0.3$ mas, $-0.8\pm 0.2$  mas)  as indicated as the center of the solid circle in the least-squares circle method \citep{2018NatCo...9.2534Y}. The estimated position of the central star is consistent with \Gaia\ EDR3 result ($-6.5\pm 0.3$ mas, $+1.5\pm 0.4$ mas) with an uncertainty of 2.7 mas on 2020 January 10.  
The differences  between the estimated position and \Gaia\ EDR3 result  ($-6.5\pm 0.3$ mas, $+1.5\pm 0.4$ mas) are 1.4 mas ($\sim$3 sigma) in east direction  and 2.3 mas ($\sim$5 sigma) in north direction on 2020 January 10.

\subsection{The expansion of the BX Cam flow traced by \hho\ masers} 
\label{sec:3Dkinematics}

\begin{figure}[t]
\epsscale{1.0}
\begin{interactive}{animation}{bxcam_h2o.mp4}
\plotone{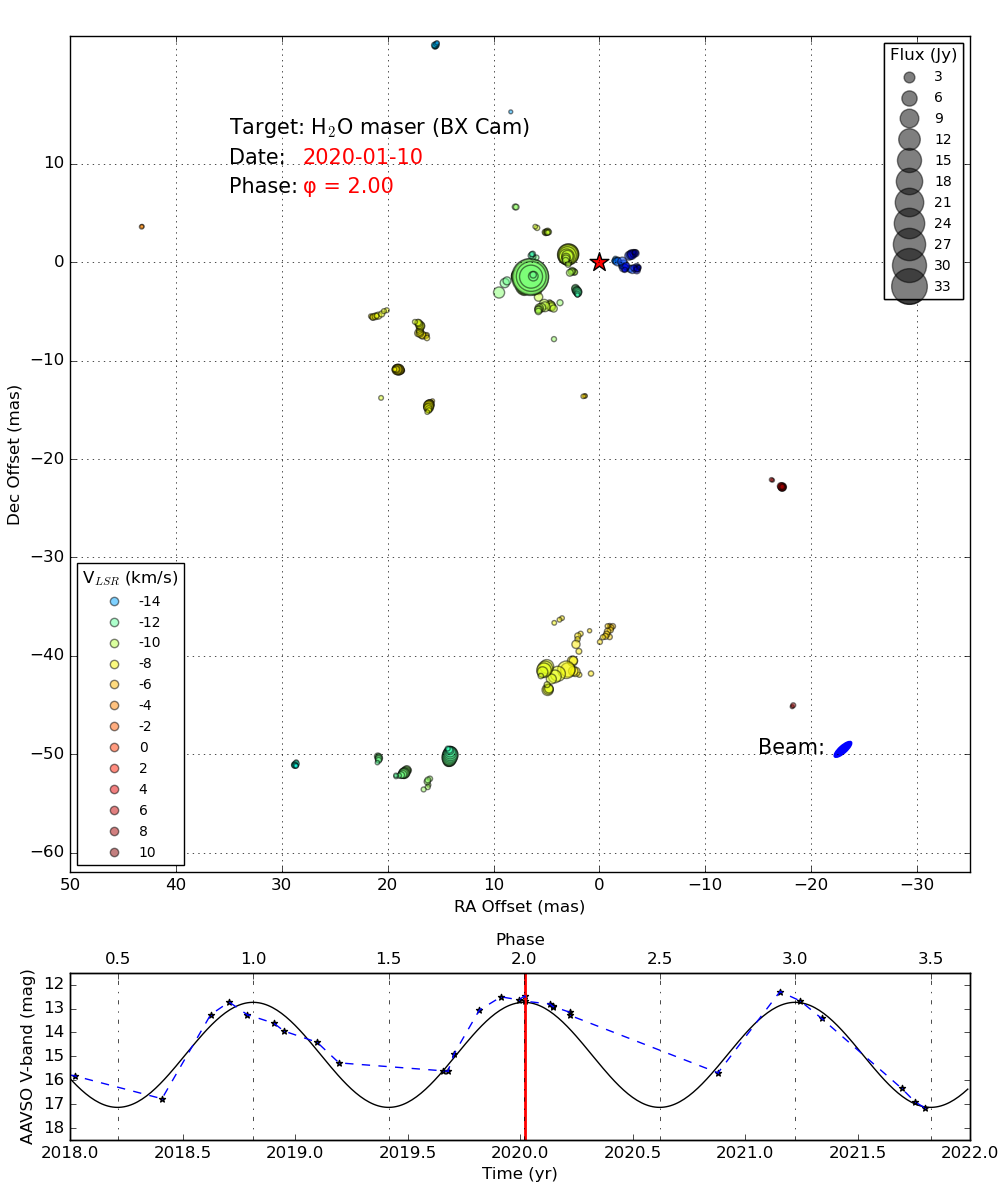}
\end{interactive}
\caption{Astrometric Animation of \hho\ masers around BX Cam. The coordinate system 
is astrometrically fixed with the star whose position is determined using \Gaia\ EDR3. 
Maser feature motions on the sky with modulation due to the annual parallax (1.79$\pm$0.08 mas) and the proper motion (14.37$\pm$0.20, $-35.46\pm 0.44$ \masy) of the referenced maser feature MF1 subtracted.
A red-filled circle denotes the central star whose position (parallax: 1.76$\pm$0.10 mas, proper motion: 14.29$\pm$0.07, $-34.63\pm 0.10$ \masy) has been determined in \Gaia\ EDR3.  The lower panel is the optical light curve of AAVSO and the red vertical line follows the phase.
An animated version of this figure is available. It covers 33-epoch maser maps from 2018 May 24 to 2021 Jun 7. The video duration is 16 seconds.
\label{fig:gaiavideo}}
\end{figure}

Figure \ref{fig:gaiavideo} displays the astrometric animation of \hho\ masers at the 33 epochs from 2018 May 24 to 2021 Jun 7.
Because of variable intervals of the observation epochs, including large blanks of the observations, the final version of the  animation will be edited with some epoch-interpolation technique \citep{2013MNRAS.433.3133G}.  Even the current version of the animation, one can recognize radial outward motions of the \hho\ maser features, helping us for noting interesting properties of the maser motions described as follows.  

\begin{figure}[t]
\epsscale{1.2}
\plotone{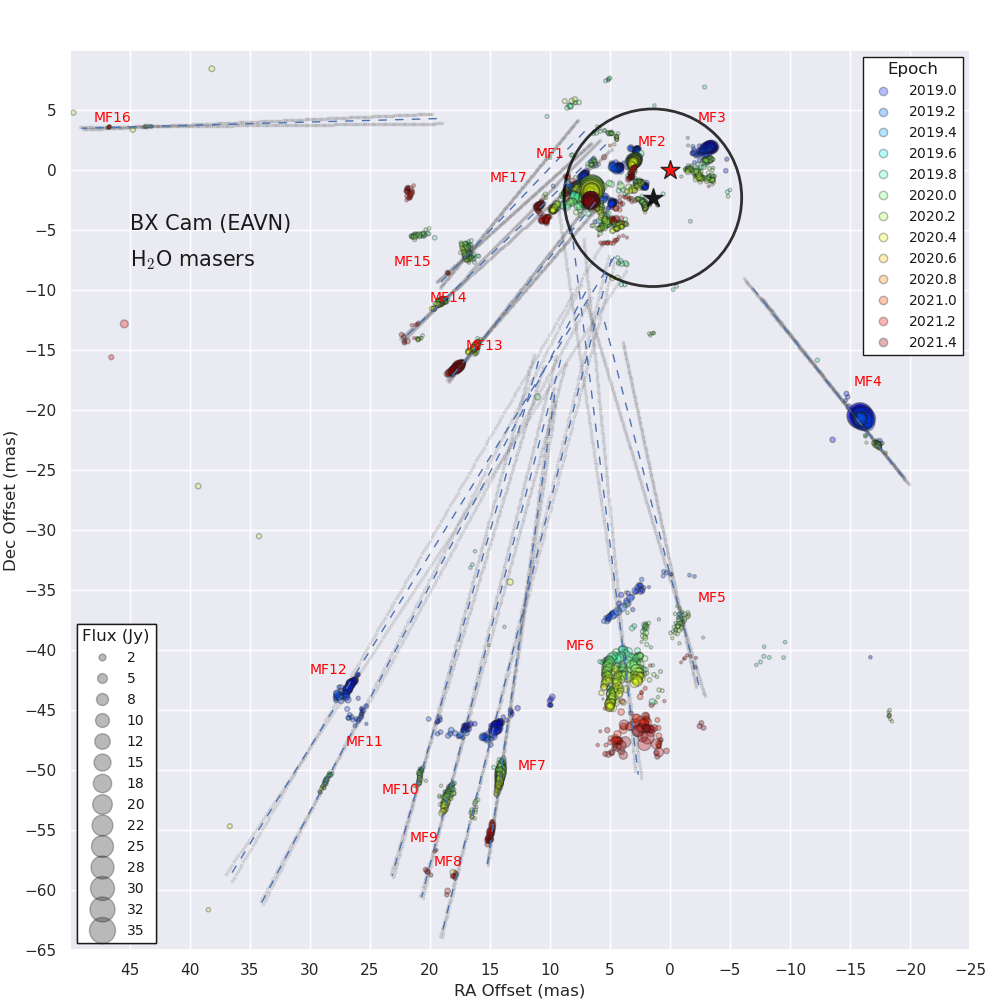}
\caption{Expansion of the BX Cam flow traced by \hho\ masers. The positional reference is \Gaia\ EDR3 (red star). Black circle denotes the fitted ring of SiO masers in Section \ref{sec:sfpr}. The different colors of the maser spots indicate different observation epochs, which are astrometrically fixed using the method in Figure \ref{fig:gaiavideo}.
An auxiliary dashed line indicates a possible trajectory of an maser feature from 2012 to 2022 at a constant velocity vector, while the nearby lines in light grey show the position error of the back-extrapolated motion vectors.
The lines of motion are drawn with the positions and proper motions of the maser feathers estimated in Table \ref{tab:vel}.
\label{fig:exp_time}}
\end{figure}

\begin{figure}[t]
\epsscale{1.2}
\plotone{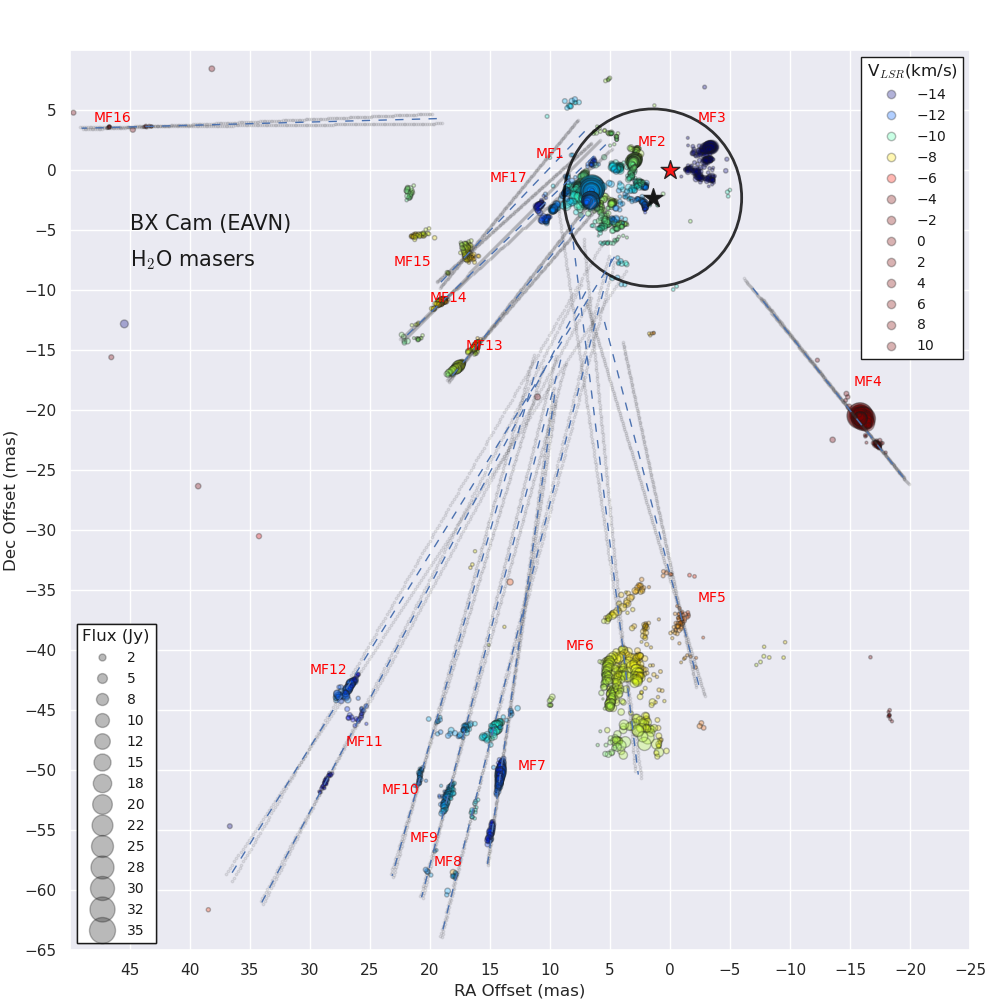}
\caption{Same as Figure \ref{fig:exp_time} but the different colors of the maser spots indicate different LSR velocities.
\label{fig:exp_vel}}
\end{figure}

Figure \ref{fig:exp_time} and \ref{fig:exp_vel} present the combined cube of the maser features synthesized with the cubes from the ``High-Res. Imaging" in Table \ref{tab:obs}.
The constant velocity proper motions and spoke-like maser features in BX Cam  ``point back" towards the central star. 
However, the point-back directions of the maser features in different groups (e.g, group MF[7,8,9,10,11,12] and group MF[13,14,15]) may converge into different areas, which may also have offsets from the location of the central star and the ring of SiO masers, as shown by auxiliary dashed lines. Thus the \hho\ maser kinematics may have origins different from the central star, if we assume that masers have the constant proper motions.

\begin{figure}[t]
\epsscale{1.2}
\plotone{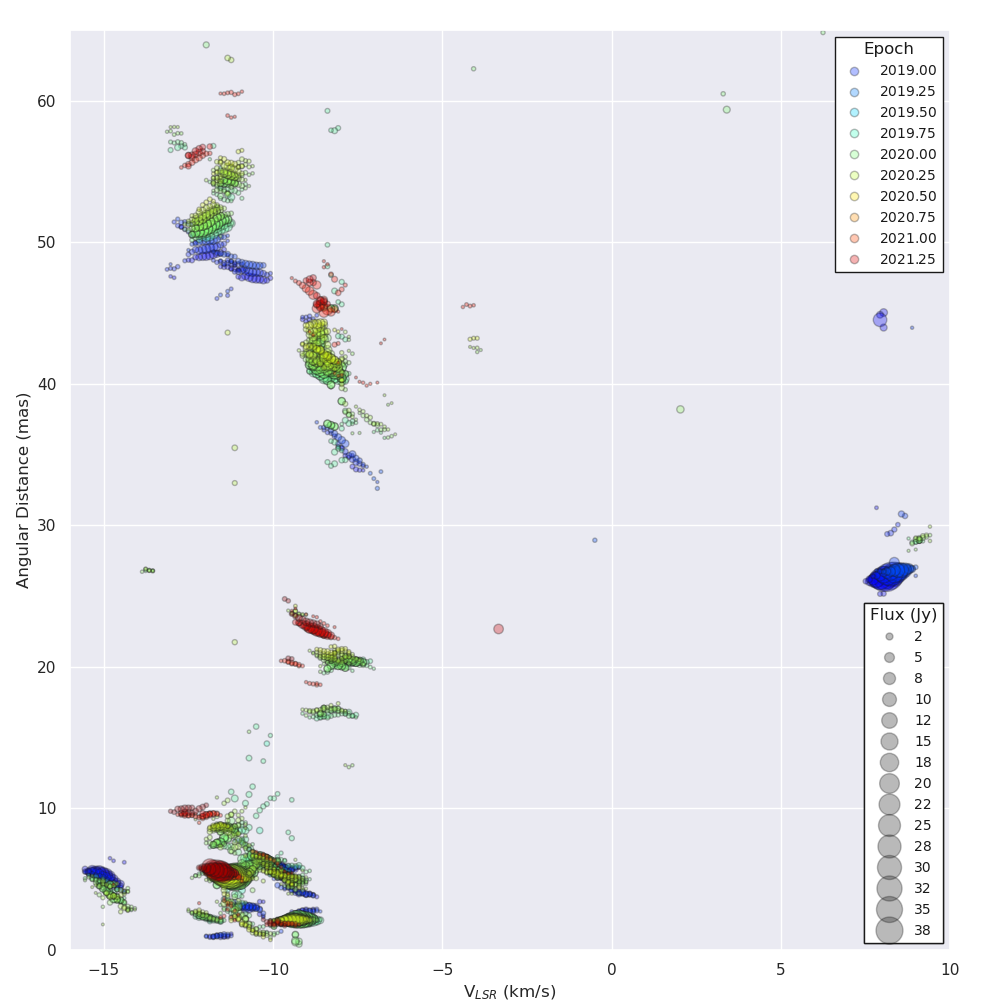}
\caption{Angular distance--LSR velocity diagram. The position of the star was used as the center of the SiO maser ring.
\label{fig_pos_vel}}
\end{figure}

The CSE of BX Cam were reported with the LSR velocity range from -5 to +5 \kms\ for SiO masers \citep{2020PASJ...72...56M},  from -16 to -4 \kms\ and from +7 to +10 \kms\ for \hho\ masers (Figure \ref{fig_pos_vel}),  from +12 to +17 \kms\ for OH masers \citep{1985A&AS...59..465S} , and up to 21 \kms\ for CO $J=1\rightarrow 0$ line \citep{1985ApJ...292..640K}, which shows a gradual acceleration from SiO maser to CO $J=1\rightarrow 0$ line. 

\begin{figure}[t]
\epsscale{1.0}
\plotone{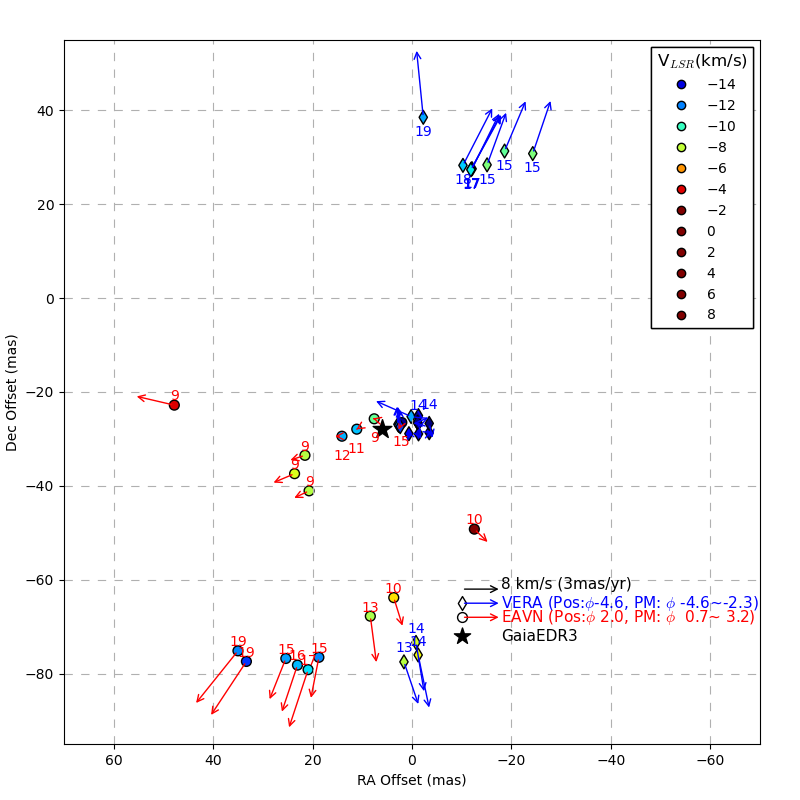}
\caption{The registered maser flows from EAVN (ESTEMA) and VERA \citep[Table 4 of][] {2020PASJ...72...56M} observed at different pulsation circles.   The registration is performed with \Gaia\ EDR3.
%in 10 February 2012 ($\phi$ -4.63).
The numbers around the maser features are the three-dimensional velocities in \kms.
\label{vera}}
\end{figure}

Figure \ref{fig_pos_vel} shows the angular distance--LSR velocity diagram of the \hho\ maser features.
To examine the variation of maser flows on long timescales, we compared the maser maps made by VERA from 2012 to 2014 \citep{2020PASJ...72...56M} and made by ESTEMA from 2018 to 2021 (Figure \ref{fig:exp_time}).
Figure \ref{vera} shows the registered map of the ESTEMA maser flows with  that of \cite{2020PASJ...72...56M}. 
Adopting a radial velocity of BX Cam as 0 \kms\ \citep{2020PASJ...72...56M}, we derived the three dimensional expanding velocities of \hho\ maser features using \VLSR\ and internal proper motions.
The maser map of \cite{2020PASJ...72...56M} showed only the blueshift dominant features with respect to the central star.
They obtained a three-dimensional velocity of 14.8$\pm$1.4 \kms\ with three collimated flows, and concluded that the \hho\ masers trace an outflow with a quite uniform velocity.
However, our result shows a wider range of 9 \kms\ (inner radii) to 19 \kms\ (outer radii) at different directions. These differences may be caused by the different observational epochs (different pulsation phases) originated from characteristics of the Mira variable star BX Cam, which traces the different expanding shells relative to \cite{2020PASJ...72...56M}.

Base on Figure 9--13, there are three main groups of \hho\ maser features. The first one (Group A: MF[1,2,3,13,14,15,17])
is associated with the inner blue-shifted maser feature closer to the central star, the second 
(Group B: MF[5,6,7,8,9,10,11,12]) 
is associated with the outer blue-shifted lobe of the outflow, and the third 
(Group C: MF4) 
is associated with the red-shifted maser feature  of the outflow. The red-shifted maser feature of the outflow (from behind the central star) is very time variable as shown in Figure \ref{fig:spec}.
The three-dimensional velocities are estimated to be 12.8$\pm$3.5 \kms\ for all maser features,  10.5$\pm$2.0 \kms\ for Group A,  15.5$\pm$2.8 \kms\ for Group B, and 10.1 \kms\ for Group C (MF4).

\begin{figure}[t]
\epsscale{0.5}
\epsscale{0.5}\plotone{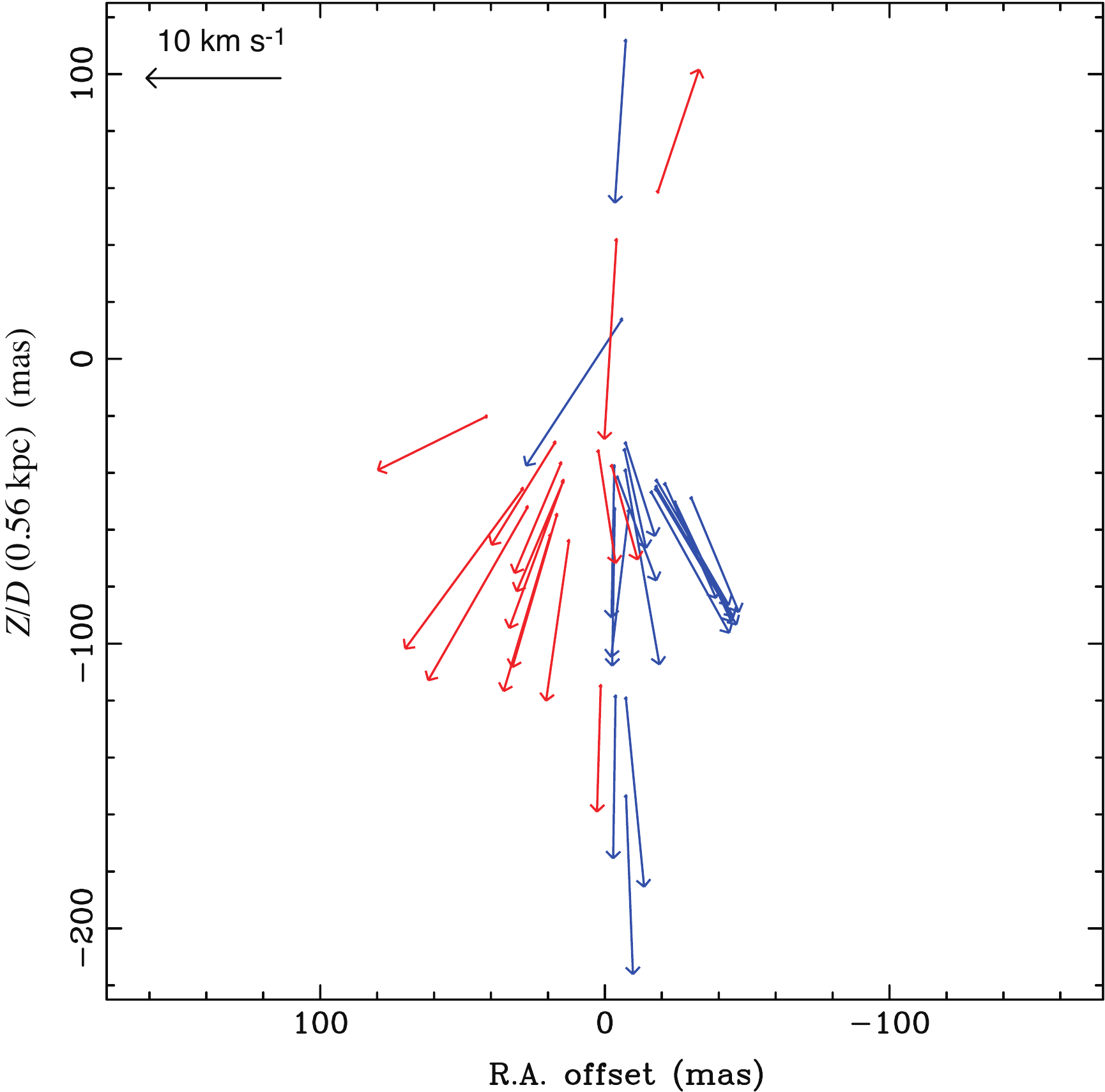}
\epsscale{0.5}\plotone{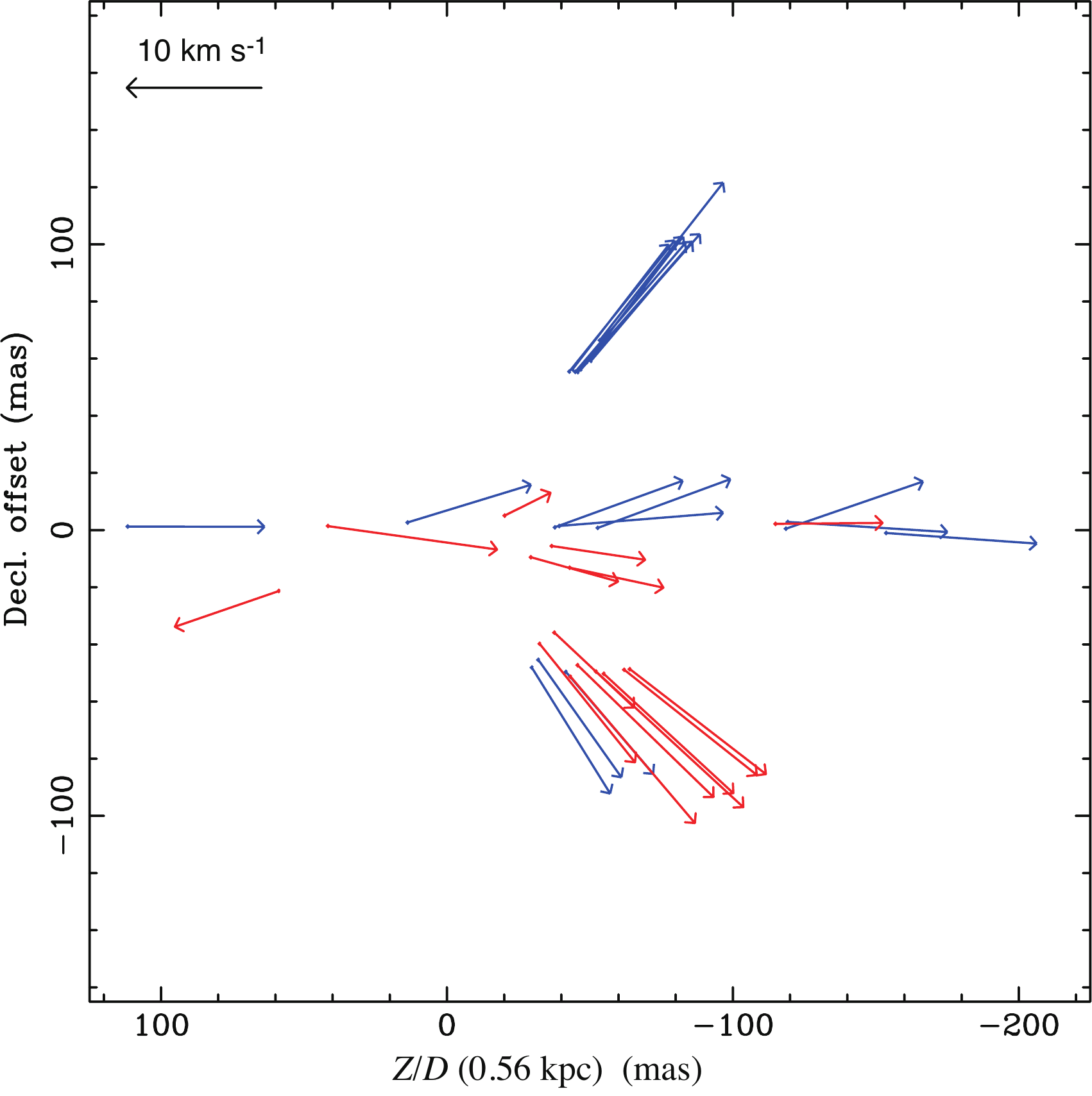}
\caption{
The Top view  (Left panel: R.A.-Distance,  XZ-plane) and the East-side view (Right panel: Distance-Dec., ZY-plane) of the estimated positions and  proper motions with respect to the outflow origin, using the method of \citet{2000ApJ...538..751I}.
The base of the arrow is pointed to the position of the maser feature. 
The direction and length of the arrow represent the direction and magnitude of the proper motion, respectively.
The red and blue arrows show the maser feature motions found in the EAVN (ESTEMA) and VERA observations \citep{2020PASJ...72...56M}, respectively.
\label{3d}}
\end{figure}

We also made a least-squares model-fitting analysis assuming a model of an outflow that is radially expanding at a constant velocity. The detail of the modeling has already been described in \cite{2012ApJ...744...23Z}. 
We adopted weights proportional to the square of the accuracy of a measured proper motion. The expansion center was 
assumed to be at the ring center of the SiO masers whose map registration was yielded in KVN SFPR (Section \ref{sec:metry}). 
The systemic velocity is adopted to be $\sim$0~\kms\ \citep{2020PASJ...72...56M}.
Thus the expansion velocity of \hho\ masers is estimated to be 13.0 $\pm$ 3.7 \kms, which is consistent with the result (14.8 $\pm$ 1.4 \kms) 
that found in the VERA observations during 2012--2014 \citep{2020PASJ...72...56M}.
Figure \ref{3d} shows the spatial distribution and velocity vectors of the maser features with respect to the outflow origin projected on the R.A.-Distance (XZ) and  Distance-Dec (ZY) planes, which was obtained after  
the model fit assuming a radially 
expanding flow \citep{2000ApJ...538..751I}. 
The outer masers 
roughly exhibit radial expansions and are 
located at similar distances from the central star at roughly a constant expansion velocity. Thus the masers seem to be associated with a spherically expanding outflow although the directions of the maser excitation are significantly biased and changed from time to time on the timescale of several years. However, there exist some big deviations from the spherical flow for
the inner masers projected closely to the central star in the sky, which may be caused by the result of enhancement of large uncertainties in the maser positions on Z-axis and some random maser motions in the modeling.

%\clearpage
\section{Discussion}

\subsection{Astrometry for CSE \hho\ masers}
\label{subsection4.1-astrometry}
Our work has evaluated the accuracy of the annual parallax measurement with CSE \hho\ masers, majority of whom  are located in the stellar neighborhood and exhibit extended brightness distributions.
The importance of astrometry is also
demonstrated for the maser animation, which is a key to register the maser images at different epochs and extract the intrinsic motions of maser features associated with the CSE. 

The astrometric accuracy of the VERA was evaluated in detail by \cite{2020PASJ...72...52N}, which achieves the parallax accuracy of 10 \uas\ in the best cases.  
Our targets were observed with a background calibrator 2\deg\ away, and at elevations of $\sim$ 50\deg. 
The single-epoch position accuracy for a given baseline can be estimated to be $\sim$80 \uas\ using the Table 1 of  \citet{2020PASJ...72...52N}, indicating an expected parallax uncertainty better than 30 \uas\ with 10 epochs extended over more than one year.
However, our derived parallax (1.79$\pm$0.08 mas) has an uncertainty more than twice the expected one, 
probably due to the variation in the maser structure (MF1 in Figure \ref{figmf})  or an irregular motion due to shocks or radiation-coupled flow as shown in the Figure \ref{fig:plxmf1}.
Nevertheless it is consistent with that previously derived with VERA (1.73$\pm$0.03 mas)  \citep{2020PASJ...72...56M} and \Gaia\ EDR3 (1.76$\pm$0.10 mas) \citep{2021A&A...649A...1G} within 1 sigma.
These parallaxes also rule out a larger parallax (4.13$\pm$0.25 mas) reported in \Gaia\ DR2 \citep{2018A&A...616A...1G}.
There is evidence that there may exist uncertainty in the \Gaia\ DR2 parallaxes for the AGB stars from their large sizes, temporal and spatial variation in the surface brightness, and/or extinction by the circumstellar dust \citep{2019ApJ...875..114X}. These factors may prevent the improvement of the parallax accuracy even in future \Gaia\ results using more measurements and better calibration of systematics.

As already shown in Section \ref{sec:metry}, the individual maser features also exhibit some deviations from linear motions. 
The smaller residuals in Figure \ref{fig:plxmf1}
support the parallax fitting including such an acceleration. However, it is obscure whether the observed acceleration really reflects a real one rather than just an illusion caused by a drift of a maser emission area in the maser gas clump. This should be tested with further analysis of the observed accelerations in the combined maser features around this star.
The residual parallaxes of all
maser features found in the EAVN maps are probably affected by their extended structures, the errors in the astrometric registration of the EAVN and VERA maps, and/or the possible non-linear motions.
Note that an additional error contribution as mentioned above may have possibilities to increase and decrease the parallax value.
Therefore, we adopted a parallax value using MF1 only (Section \ref{sec:plx}) and give a conservative uncertainty that would cover all of the possible errors (Section \ref{sec:plx-other}).

The stellar position in the \hho\ maser distribution was estimated to be at the ring center of the SiO masers whose locations are determined in the KVN astrometry with the SFPR technique at a single epoch. The SiO maser ring structure will be further clarified with multiple maser lines \citep{2018NatCo...9.2534Y}. Even the present results, one can clearly see an offset of the expansion center of the \hho\ maser flow from the center of the SiO maser ring as shown in  Figure \ref{fig:exp_time}. This may be caused by either the uncertainty of the stellar proper motion measured in \Gaia\ EDR3 and adopted in the maser map registration, or a possible deviation of the outflow from a radial expansion.

\subsection{CSE spatio-kinematics: asymmetric maser distribution and its  evolution}

The asymmetry/inhomogeneity of the spatial distribution is evident throughout the three-dimensional maser kinematics of \hho\ masers that we have revealed. The maser outflows are dominated by the blue-shifted \hho\ maser features (dominant blue-shifted outflow of bipolar outflow).
Compared to the maser features detected with VERA from 2012 to 2014 \citep{2020PASJ...72...56M}, we have detected the central and southern components, and also new red-shifted components. The northern components previously mapped by  \cite{2020PASJ...72...56M} have become so faint that they are detected only on KVN baselines, at a few epochs.

Although the velocity field of the masers is generally consistent with 
the expanding shell model, the expansion is not strictly isotropic. For BX Cam, the distribution of the maser features is limited in a thick-shell region (Section \ref{sec:3Dkinematics}), which has inner and outer expansion velocities of 9 \kms\ and 19 \kms\ at radii of $<$20 mas and 60 mas, respectively.  There also exist  gradual drifts in radial velocities and possible accelerations/decelerations in the proper motion.
Here, we suppose a constant/gradual radial acceleration in a CSE (e.g. \citealt{2012IAUS..287..199R}) for BX Cam. Adopting the distance to the star $D=$560~pc, one can calculate a travel time from the inner and outer radii of the expanding maser shell to be $t_{\rm travel}\simeq$7.6~yr and an acceleration to be $a\simeq$1.3~\kms\ yr$^{-1}$, corresponding to $\dot{\mu}\simeq$0.5~mas~yr$^{-2}$.
This suggests that observed drifts of LSR velocities can be explained by this kind of acceleration while the accelerations/decelerations seen in the maser proper motions (up to 1.8~mas yr$^{-2}$) cannot. Because the accuracy of the observed accelerations/deceleration is currently insufficient for further astrophysical interpretation, we reserve further statistical analysis for a separate paper. 
Nevertheless, the existence of such accelerations or deviations from constant velocity motion, 
may be detectable in the systematic residuals after the parallax and proper motion fitting.

The expansion velocity and the projected distance to the central star of a mass-losing flow are seem to be time-dependent. The flows found in the ESTEMA data were compared with that in the VERA observations from 2012 to 2014 in Figure \ref{vera}.  Different shells and distributions were detected at different times and with different arrays. 

Temporal variation in the morphology of individual maser features is also determined.
The shapes of the maser structures are generally correlated with the direction of expansion. Figure \ref{figmf} also shows the complex variation of the interior structure, such as interior rotation, expansion, sub-structure, and change of direction. This might be due to any or all of: bulk rotation, internal turbulence, or a pattern speed due to shock wave propagation \citep{2012A&A...546A..16R,2020AdSpR..65..780R}.

\subsection{Variability of the \texorpdfstring{\hho}{H2O} maser }

A correlation between the 22\,GHz \hho\ maser and optical light curves with a time lag has been reported in AGB stars \citep{1993LNP...412..271B}. BX cam is in also such a case, and the measured time lag between the integrated water maser and optical light curves is about $\sim$25\,d.
However, several bright masers often dominate the total integrated intensity, and thus it is hard to depict the variation of individual maser features or spots without spatially resolved maser maps. Our VLBI images allow us to trace individual maser features (shown in Figure \ref{fig:timemaser}) and groups of maser features (shown in Figure \ref{fig:submaser}). We find that the light curves of the individual maser features or groups have different time lags with the optical light curve.
We note that \hho\ masers are pumped by collision rather than radiation field, and hence other factors should be considered to interpret the correlation between maser and optical light curves, such as shock waves caused by stellar pulsation or giant planets, revolving the central star \citep{1988ApJ...330..986S,2001Icar..151..130R,2012A&A...546A..16R}. 

\section{SUMMARY}
\label{sec:sum}

We have measured the trigonometric parallaxes, proper motions, drifts in LSR velocities, and the possible accelerations/decelerations
of the \hho\ masers around the Mira variable BX Cam using EAVN observations.  
Our parallax is 1.79 $\pm$ 0.08 mas, which is consistent with \Gaia\ EDR3 and previously measured VLBI parallaxes within their joint
uncertainties.
The stellar position with respect to the \hho\ masers is registered by a ring of SiO masers using the KVN SFPR astrometry, which is consistent with that of \Gaia\ EDR3 within 3 mas. 
There are asymmetries/inhomogeneities  in both the spatial and velocity distributions of the \hho\ masers
although their motions are roughly modeled with a spherically expanding outflow. In fact, the maser locations have been dominant in a blue-shifted outflow with respect to the central star.
The 3D \hho\ maser kinematics indicates that the circumstellar envelope is expanding at a velocity of $13\pm4$ \kms. A time lag of about 25\,d is also measured between the integrated water maser and optical light curves.
More detailed maser kinematics for the CSE around BX Cam will appear in separate papers, covering 
multiple maser-line movies with SiO masers and the further discussion on the possibility of the maser clump acceleration.  

\acknowledgments
We acknowledge with thanks the variable star observations from the AAVSO International Database contributed by observers worldwide and used in this research. We are grateful to all staff members in EAVN who helped to operate the array and to correlate the data. Particularly, the interferometer model (IM) files are provided by Mr. DongKyu Jung, and the delay re-calculation tables are made by Dr. Takumi Nagayama and Dr. Kazuya Hachisuka.
The KVN is a facility operated by Korea Astronomy and Space Science Institute (KASI) and VERA is a facility operated by the National Astronomical Observatory of Japan (NAOJ) in collaboration with associated universities in Japan. Tianma 65-m telescope is operated by Shanghai Astronomical Observatory (SHAO). Nanshan 26-m telescop is operated by Xinjiang Astronomical Observatory (XAO). 
Takahagi 32- m telescope is operated by NAOJ and Ibaraki University and partially supported by the Inter-university collaborative project ``Japanese VLBI Network (JVN)'' of NAOJ.
HI and MO was supported by JSPS KAKENHI JP16H02167. 
LC is supported by the CAS ``Light of West China'' Program (Grant No. 2021-XBQNXZ-005) and the National Natural Science Foundation of China (Grant No. U2031212).
AMS was supported by the Russian Ministry of Science and Higher Education, No. FEUZ-2020-0030.
BZ was supported by the National Natural Science Foundation of China (Grant No. U2031212 and U1831136),
and Shanghai Astronomical Observatory, Chinese Academy of Sciences (Grant No. N2020-06-19-005). 

\vspace{5mm}
\facilities{EAVN}
\software{AIPS \citep{2003ASSL..285..109G}, ParselTongue \citep{2006ASPC..351..497K}, Astropy \citep{2013A&A...558A..33A}}

\bibliography{ref}{}
\bibliographystyle{aasjournal}

\appendix
\section{Temporal variation of maser features in the EAVN images}
\label{Appendix-maser-structure-variation}
The spatial and velocity variation of maser features at different epochs 
were investigated with the ``High-Res. Imaging" in Table \ref{tab:obs}.
We used MF1 as the position reference, which was the brightest and compact maser feature during the period from $\phi$ = 1.88 to 3.17, to register the EAVN maps at different epochs.
The registration accuracy can be expressed as $\sigma_{registration} = \sqrt{ {\sigma_{{thermal}}}^2 + {\sigma_{{structure}}}^2}$, which includes the thermal noise ($\sigma_{thermal}$) in the interferometric images determined by an elliptical Gaussian fitting with the AIPS task ``SAD'', includes error from maser structure ($\sigma_{structure}$) estimated by the blended peak components as shown in Figure \ref{fig:timemaser}, and doesn't include  other calibration effects for MF1.
When MF1 was not available, we used MF2 instead, and corrected the position offsets using the bona-fide astrometric measurements in Table \ref{tab:fitall}. 
The registration error is typically $\sim$ 0.1 mas for using MF1 cases ($\sigma_{thermal} \simeq$ 0.02 mas, $\sigma_{structure} \simeq$ 0.05 mas) and $\sim$ 0.2 mas for using MF2 cases ($\sigma_{thermal} \simeq$ 0.07 mas, $\sigma_{structure} \simeq$ 0.15 mas). These registration errors are used as the prior errors in the astrometric fitting.

Figure \ref{fig:timemaser} shows the spatial and velocity variation of maser features. 
Here we found the following four points. 
Firstly, the shapes of the maser structures generally correlate with the direction of expansion. Figure \ref{figmf} also shows the complex variation of the interior structure, such as 
rotation of MF1, expansion on the shape of MF2, sub-structure flashed on MF4, direction changed on MF3 and MF13.
Secondly, as shown in Figure \ref{fig:timemaser} and Table \ref{tab:vel}, the relative proper motions are not necessarily consistent between different Cycles for some maser features. The absolute proper motions  and possible accelerations are discussed in section \ref{sec:metry}. 
Thirdly, as shown in Figure \ref{fig:timemaser} and Table \ref{tab:vel}, almost all of the maser features have the 
radial velocity drifts.  
Generally, the red-shifted maser features are further red-shifted, while the blue-shifted features are further blue-shifted. 
The systematic 
radial velocity drifts
of individual maser features were found with amplitudes of $<$ 1.3 \kmsyr, which is consistent with the case of RT Vir \citep{2003ApJ...590..460I}.  
The 
radial velocity drifts
are not consistent between different Cycles,  which shows the gradual behaviors related to the pulsation phases.
For instance, the mean of the drifts in Cycle 2 ($\phi\ 1.89\sim 2.29$) is larger than that over all the cycles. 
Finally, the brightest flux density of each maser feature appears at different Cycles in Figure \ref{fig:timemaser}, such as MF[3,4,8,12] appear in Cycle 1 ($\phi\ 1.00\sim 1.25$),  MF13 appears in Cycle 3 ($\phi\ 2.89\sim 3.17$) and others appear in Cycle 2 ($\phi\ 1.89\sim 2.29$). In Cycle 2 ($\phi\ 1.89\sim 2.29$), most of maser features are detected and they all peak on 2020 January 10 ($\phi$ = 2.00), except for MF8 and MF16 that are located far from the central star.

As shown in Figure \ref{fig:submaser}, we have further investigated the coherent variation among clusters of maser features located at different projection distances from the central star.
In Cycle 2 ($\phi\ 1.89\sim 2.29$), there is no noticeable difference in the phase lags of flux variation between different regions. In Cycle 1 ($\phi\ 1.00\sim 1.25$) \& 3 ($\phi\ 2.89\sim 3.17$), the peak flux densities appear at different time for different regions, with no obvious systematic behavior. 

\begin{figure}[ht!]
\epsscale{0.57}
\plotone{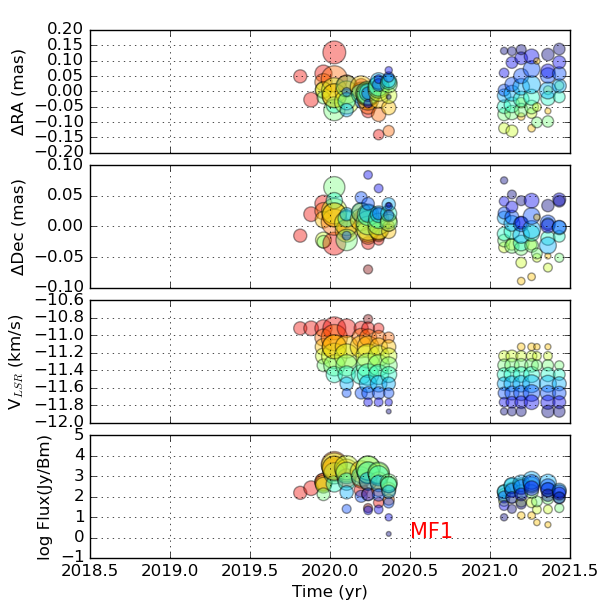}
\plotone{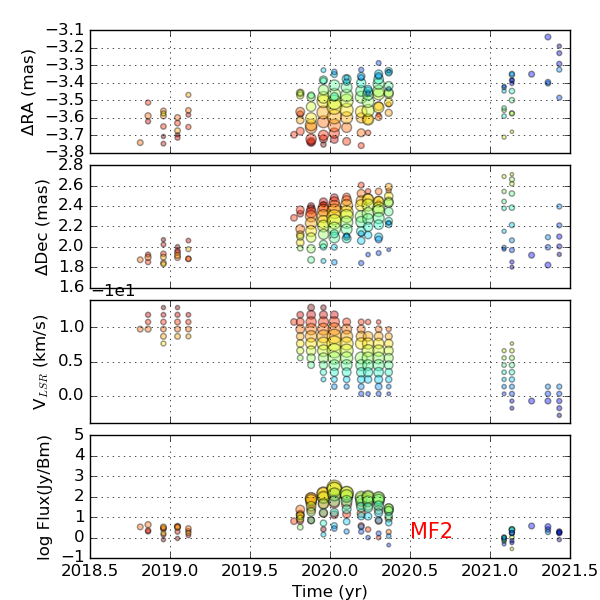}
\plotone{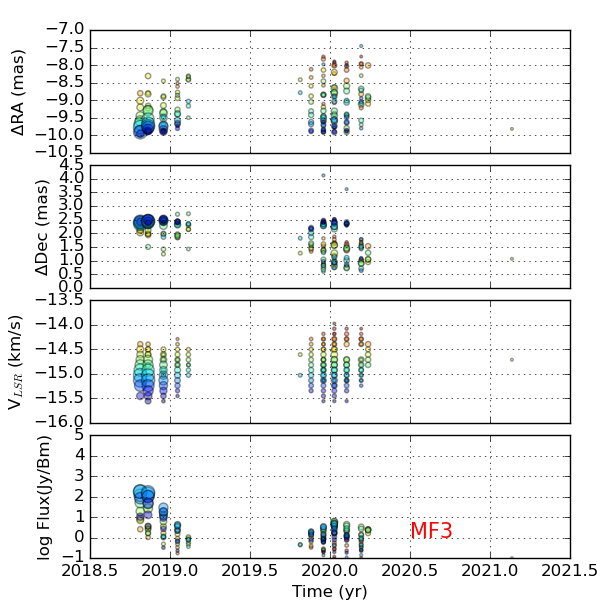}
\plotone{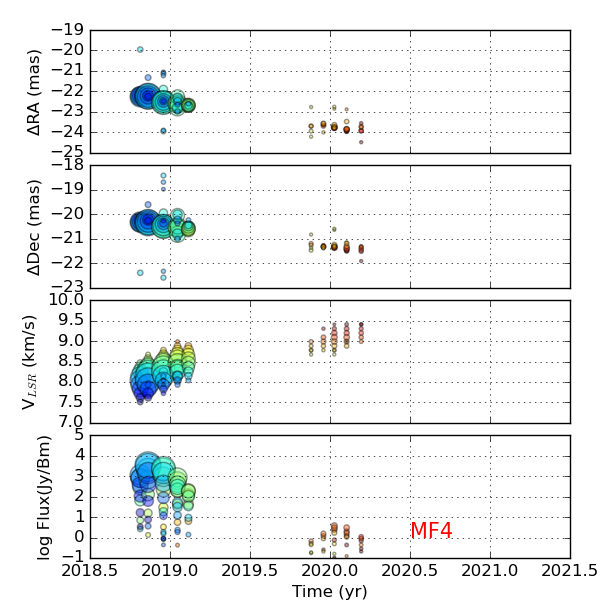}
\caption{Temporal variation of the internal maser features. The size of the circle indicates the logarithm of the flux density of the maser spot (see the fourth sub-panel for the each maser feature as a legend). The color of the circle indicates the LSR velocity of the spot (see the third sub-panel). 
\label{fig:timemaser}}
\end{figure}

\begin{figure}[t]
\epsscale{0.57}
\plotone{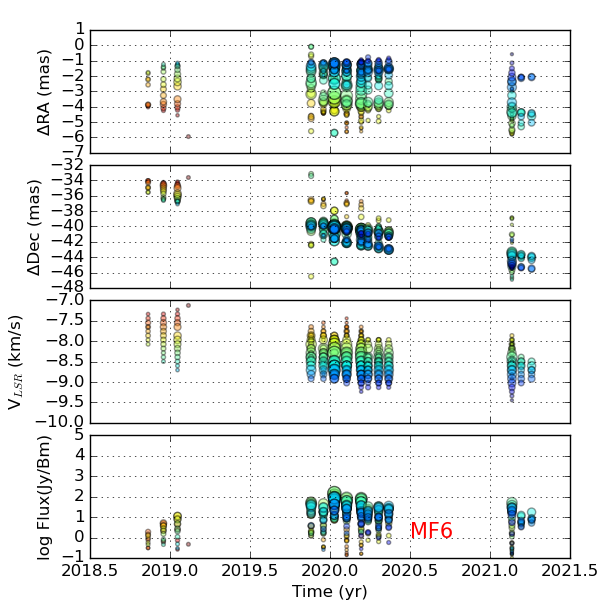}
\plotone{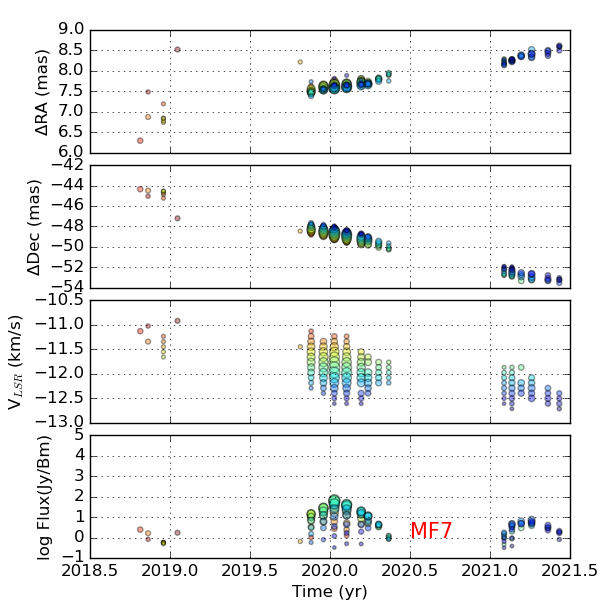}
\plotone{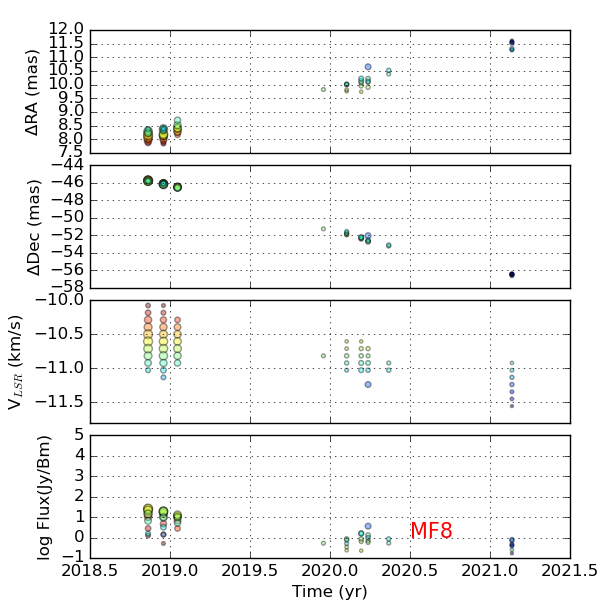}
\plotone{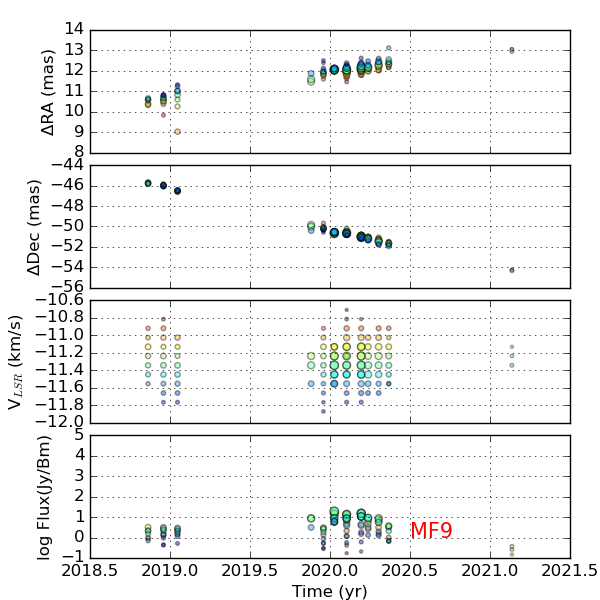}
\addtocounter{figure}{-1}
\caption{--- continued.}
\end{figure}

\begin{figure}[t]
\epsscale{0.57}
\plotone{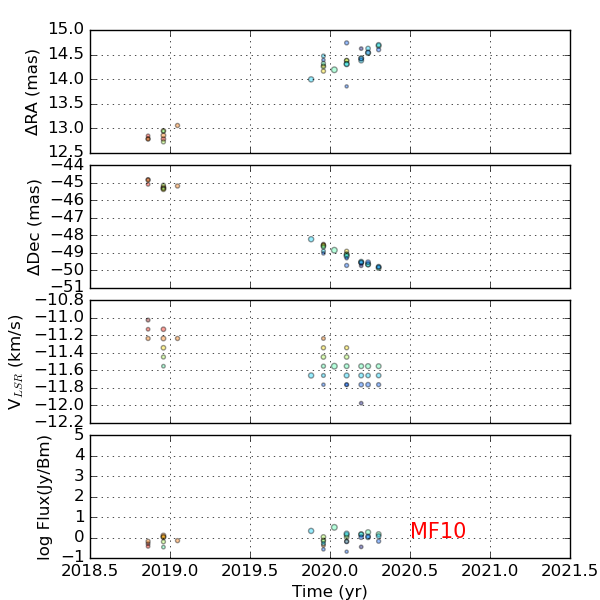}
\plotone{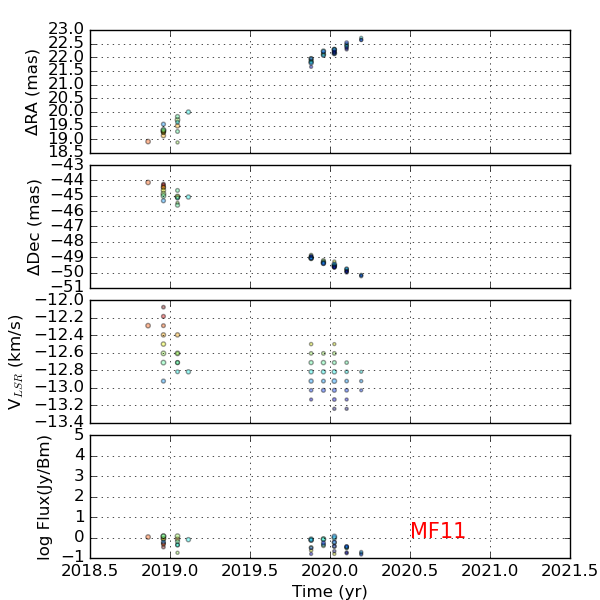}
\plotone{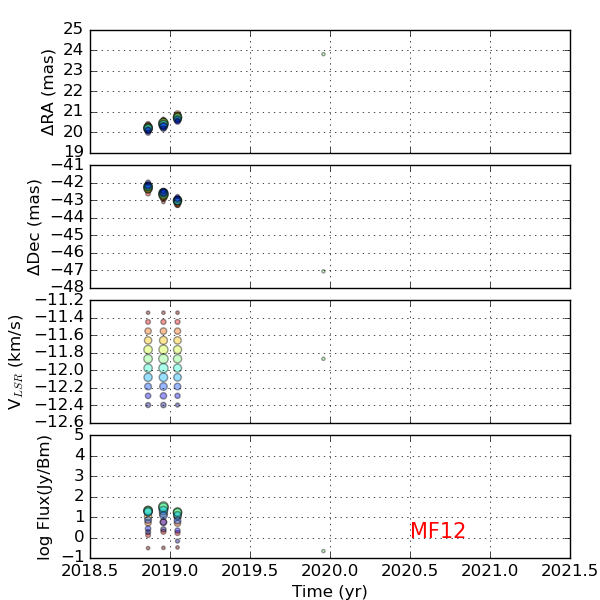}
\plotone{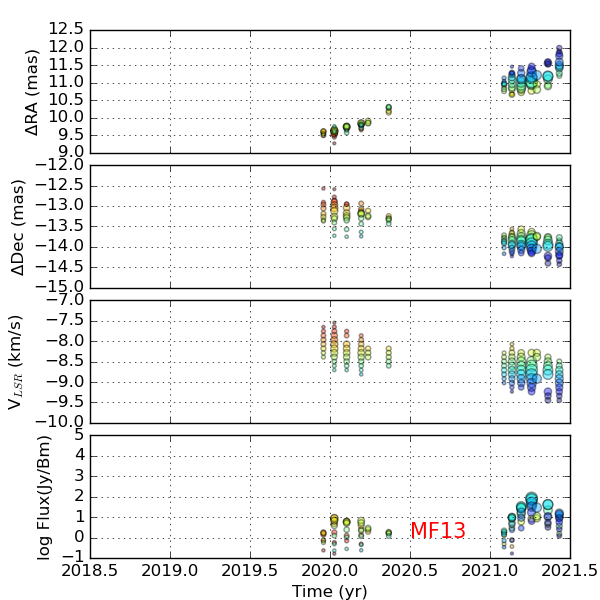}
\addtocounter{figure}{-1}
\caption{--- continued.}
\end{figure}

\begin{figure}[t]
\epsscale{0.57}
\plotone{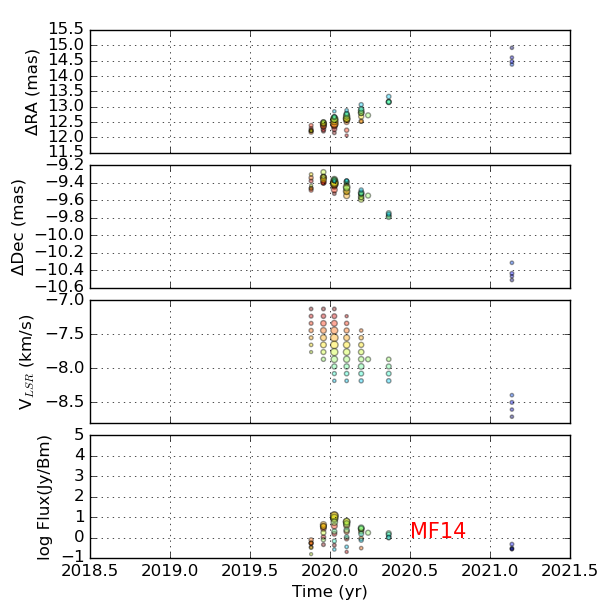}
\plotone{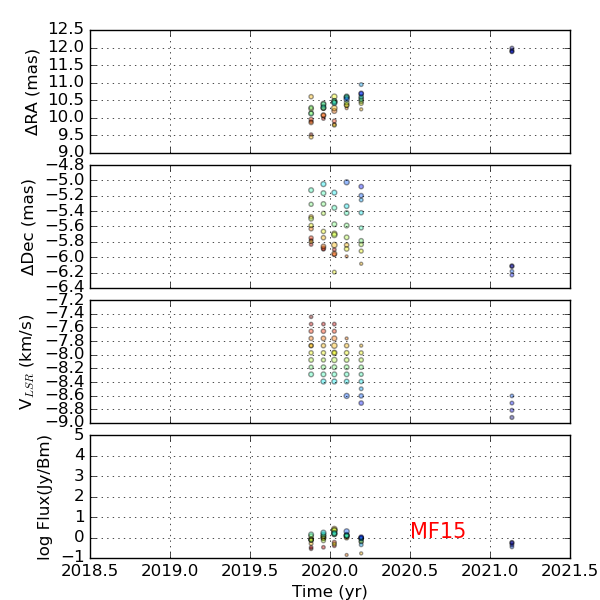}
\plotone{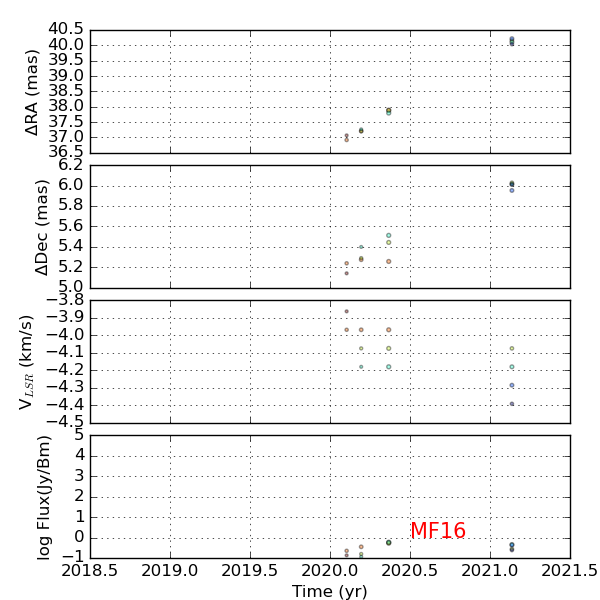}
\plotone{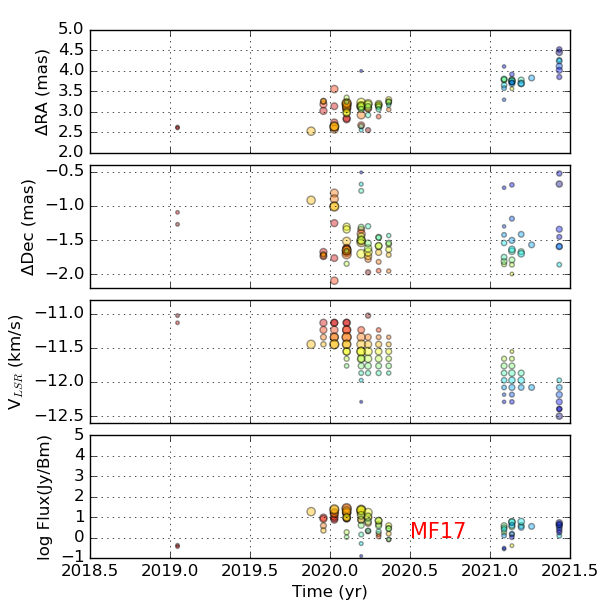}
\addtocounter{figure}{-1}
\caption{--- continued.}
\end{figure}

\begin{figure}[t]
\epsscale{1.2}
\plotone{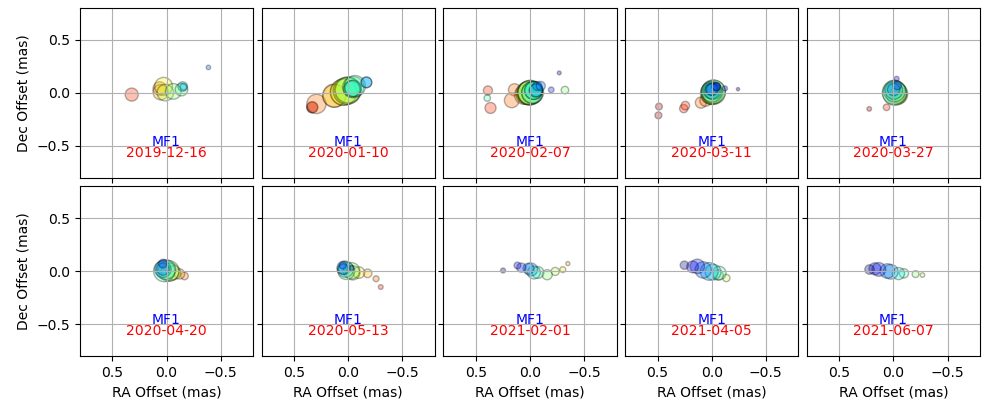}
\plotone{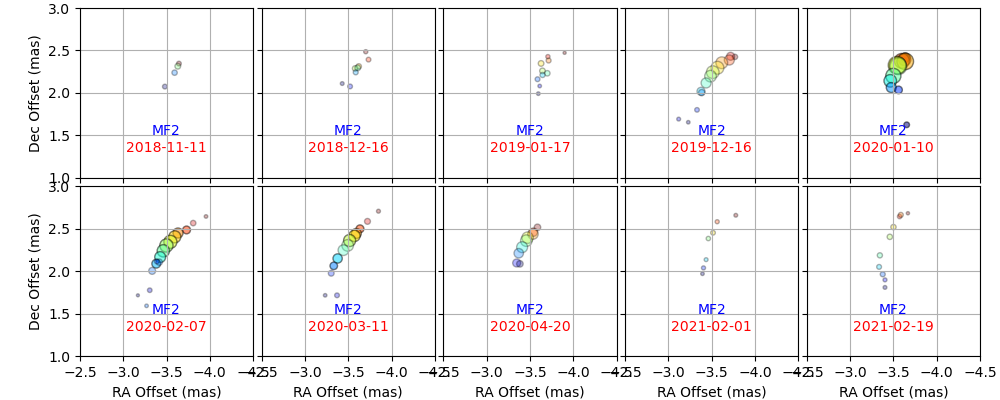}
\plotone{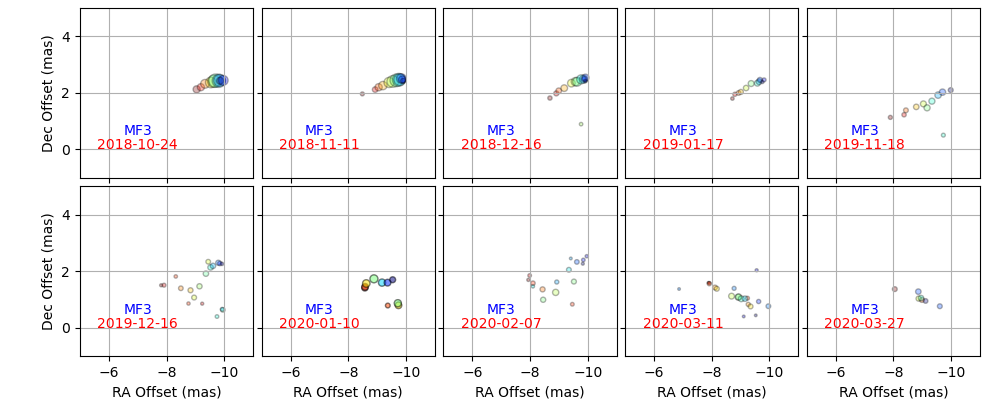}
\caption{The morphology of individual maser features. The color and size of maser spots are the same to Fig \ref{fig:timemaser}.
\label{figmf}}
\end{figure}

\begin{figure}[t]
\epsscale{1.2}
\plotone{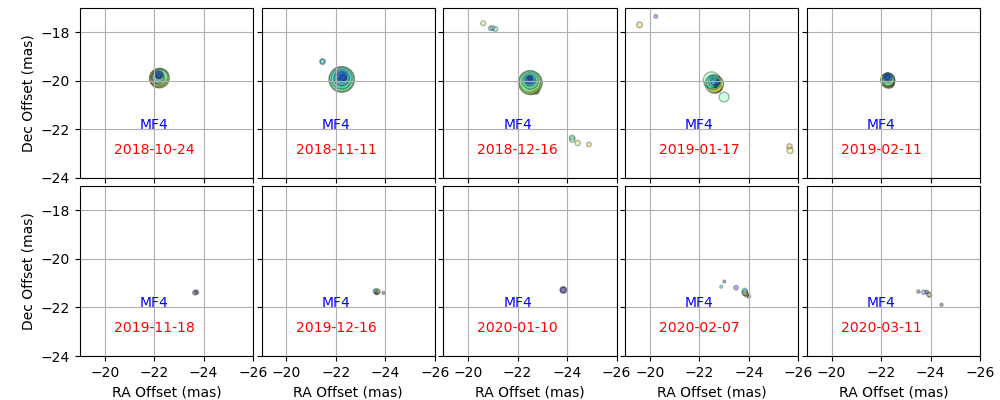}
\plotone{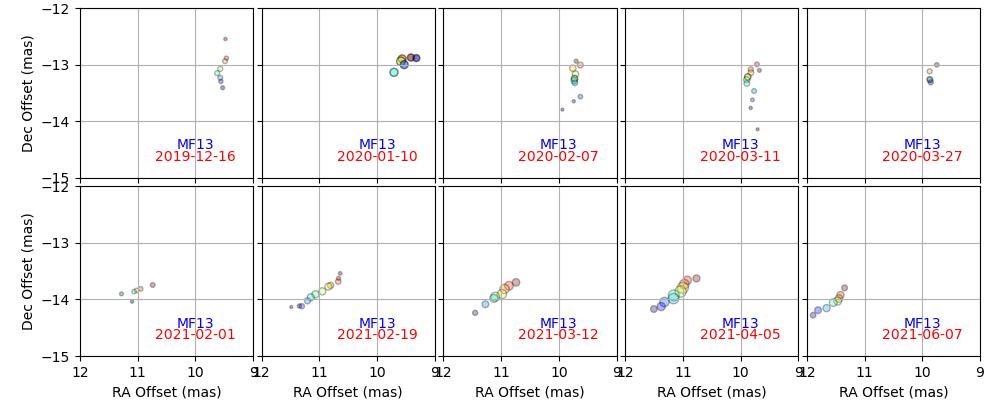}
\addtocounter{figure}{-1}
\caption{--- continued.}
\end{figure}

\end{CJK*}
\end{document}